\documentclass[aps,prb,preprint,showpacs,floatfix]{revtex4-1}

\usepackage{longtable}

\usepackage[final]{graphicx}% Include figure files
\usepackage{dcolumn}% Align table columns on decimal point
\usepackage{bm}% bold math
\usepackage{braket}
\usepackage{xcolor}

\newcommand{\mm}[1]     {\ifmmode {#1} \else{}${#1}$\fi}
\newcommand{\mmm}[1]    {\ifmmode{}#1 \else{}${#1}$\fi}
\newcommand{\beq}[1]{\begin{equation}\label{#1}}
\newcommand{\eeq}{\end{equation}}

\def \acuos{\mm{\rm A_{2}Cu_{3}O(SO_4)_3}}

\def\av#1{\mm{\left<{#1}\right>}}
\def\vec#1{\mm{{\rm\bm{{\mathrm#1}}}}}

\def\P6m{\mm{P{6}/m}}

\def\figsiz{\columnwidth}

\begin{document}


\title{\large Long range 3D magnetic structures of the spin $\bm S$=1  hexamer cluster fedotovite-like $\bm\acuos$ (A$_2$=K$_2$, NaK, Na$_2$): a neutron diffraction study }

\author{V.~Yu.~Pomjakushin}
\affiliation{Laboratory for Neutron Scattering and Imaging LNS, Paul Scherrer
Institut, CH-5232 Villigen PSI, Switzerland}

\author{A.~Podlesnyak}
\affiliation{Neutron Scattering Division, Oak Ridge National Laboratory, Oak Ridge, Tennessee 37831, USA}

\author{A.~Furrer}
\affiliation{Laboratory for Neutron Scattering and Imaging LNS, Paul Scherrer
Institut, CH- 5232 Villigen PSI, Switzerland}

\author{E.~V.~Pomjakushina}
\affiliation{Laboratory for Multiscale Materials Experiments LMX, Paul Scherrer Institut, CH-5232 Villigen PSI, Switzerland}

\date{\today}

\begin{abstract}
The crystal and magnetic structures of the spin $S$=1  hexamer cluster fedotovite-like $\acuos$ (A$_2$=K$_2$, NaK, Na$_2$) were studied by neutron powder diffraction at temperatures 1.6-290 K. The crystal structures in all compounds are well refined in the monoclinic space group C2/c. The basic magnetic units of the compounds are copper hexamers which are coupled by weak superexchange interactions giving rise to three-dimensional long-range magnetic order below $3.0<T_N<4.7$~K. We have found that for A$_2$=K$_2$ and NaK the propagation vector of the magnetic structure is $\bm k$=[0,0,0], and the coupling of the Cu hexamers is ferromagnetic FM along the $ab$ diagonal and antiferromagnetic along the $bc$-diagonal. In contrast, for A=Na the propagation vector is $\bm k$=[0,1,0], and the Cu hexamers are coupled antiferromagnetically AFM along the $ab$ diagonal. The hexamers are formed by three Cu-pairs arranged along the $b$-axis. The calculated spin expectation values \av{s} for the simplest symmetric spin Hamiltonian (obtained from inelastic neutron spectroscopy) of the isolated hexamers in the mean field, amounted to \av{s}=3/8 for side spins Cu1 and Cu2 and  \av{s}=1/4 for Cu3 in the middle. The Cu spins are FM coupled in pairs and AFM between neighboring pairs. The experimental magnetic moments of the Cu$^{2+}$ ions turn out to be not completely collinear due to spin frustrations within the weak interhexamer interactions. The sizes of magnetic moments of Cu in the hexamers determined from the diffraction data are in fair agreement with the calculated values.

\end{abstract}


\maketitle

% body of paper here
%\pagebreak

\section{Introduction}

%% Albert
Quantum magnetism and related spin frustration phenomena are relevant topics in the present surge of interest in condensed-matter physics. Of particular importance are studies of spin $S$=1/2 systems, in which spin frustration is often present due to both quantum spin fluctuations and the lack of anisotropy. Such phenomena have been observed in many compounds and particularly in compounds built up of magnetic clusters which interact as single magnetic units in the three-dimensional lattice \cite{A1,A2}. The magnetic properties are governed by strong intra-cluster and weak inter-cluster interactions, so that the interplay of different energy scales and dimensionalities can lead to unusual phenomena. More recently, an increasing number of investigated compounds benefited from observations on naturally occurring minerals \cite{A2,A3}, thereby bridging mineralogy with condensed-matter physics.
Here we focus on the minerals \acuos\ (puninite A=Na, euchlorine A$_2$=NaK, fedotovite A$_2$=K$_2$) which have been studied in the past few years by a variety of both experimental and theoretical methods \cite{A4,A5,Furrer2018,Masa2019,Furrer2020,Tsirlin,Furrer2021}. The compounds are built up of edge-shared tetrahedral spin clusters consisting of three pairs of Cu$^{2+}$ ions with spin $S$=1/2 as shown in Fig. \ref{hexamer} forming quasi-isolated hexamer. The intra-cluster interactions lead to a triplet ground-state $S$=1 for the copper hexamers \cite{A4,Furrer2018,Furrer2020}, which are weakly coupled giving rise to long-range magnetic order below $T_N$=3.4~K (A=Na), 4.7~K (A$_2$=NaK), and 3.0~K (A=K) \cite{Tsirlin}, thereby challenging the earlier description of the title compounds in terms of Haldane spin chains \cite{A3}. Later studies gave evidence for two-dimensional superexchange interactions within the (b,c)-plane \cite{Tsirlin,Furrer2021}. So far detailed information on the magnetic structures is missing, except from a neutron diffraction study performed for A=K, where the solution was found in Shubnikov group C2'/c with the moments aligned in the $ac$ plane~\cite{Masa2019}. 
We would like to note that the magnetic neutron diffraction experiments are very difficult in this system because of the small magnetic moments of Cu$^{2+}$ spin $S$=1/2 and very large natural background due to the fact that the magnetic intensities are positioned on top of huge nuclear Bragg peaks. The nuclear peaks are contributed by 168 atoms, whereas the magnetic peaks are formed only by 24 Cu spins per unit cell. In the Ref~\cite{A4}  
no magnetic Bragg peaks were detected in the neutron powder diffraction experiment down to 1.6~K in the sample with A=K. Hase et al~\cite{Masa2019} have successfully found the solution for the magnetic structure for A=K from relatively low number of diffraction peaks with only one statistically relevant magnetic reflection. The non zero magnetic moments were found only at the side Cu-pairs Cu1 and Cu2 in the hexamer.

The present work reports on the magnetic structures determined from neutron diffraction experiments carried out below and above $T_N$. We find that the compounds with A$_2$=K$_2$ and NaK are defined by the magnetic Shubnikov group $C2'/c$. The propagation vector is $\bm k$=[0,0,0], and the coupling of the Cu hexamers is ferromagnetic along the $ab$ diagonal. In contrast, for A=Na the magnetic Shubnikov group is C\_P2'/c (OG symbol~\cite{MSG}) with propagation vector $\bm k$=[0,1,0], and the Cu hexamers are coupled antiferromagnetically along the $ab$ diagonal. All Cu atoms in hexamers have non zero magnetic moment values. Our results are discussed in terms of the intra-hexamer and inter-hexamer exchange parameters available for the title compounds \cite{Tsirlin,Furrer2021}. Spin frustration turns out to be evident, which is actually expected from the empirical quantity $|T_{cw}/T_N|>>1$~\cite{A11}, where $T_{cw}$ is the Curie-Weiss temperature derived from magnetic susceptibility measurements \cite{Tsirlin}. The experimental observation of the 3D-long range magnetic order seems forbids emergence of the cluster based Haldane state in this system.    

%% END Albert

% Our previous papers on this subject:

%2018 Phys. Rev. B 98, 180410(R) Spin triplet ground-state in the copper hexamer compounds 0A2Cu3O(SO4)3(A=Na,K)
%% Table S1 contains table with structure parameters for Na2 K$_2$
%
%Spin-coupling topology in the copper hexamer compounds A2Cu3O(SO4)3 (A=Na, K) 
%PHYSICAL REVIEW B 101, 224417 (2020)
%
%PHYSICAL REVIEW B 104, L220401 (2021)Imbalanced spin coupling in the copper hexamer compounds A2Cu3O(SO4)3 (A2 = Na2, NaK, K$_2$)

% Table S2 contains table with structure parameters for NaK

%%%Quantum magnetism and related spin-frustration phenom- ena are an active field in present condensed-matter research, which has been largely inspired by the discovery of natu- rally occurring minerals [1]. The minerals A2Cu3O(SO4)3 (A2 = Na2, NaK, K2) are novel candidates which may ad- vance this field. They are of particular interest, since they are composed of molecularlike copper hexamers with strong intrahexamer and weak in
%%% 
%%%PRB,Tsirlin2020 - TN between 3, 3.4, 4.7 for K. Na, NaK Better to say below 6K
%%%PRL, Fujihala2018 TN 4K for K.
%%%Masa2019 TN = 3.1K

\section{Samples synthesis and experimental details}
\label{exp}

Polycrystalline samples of \acuos\ (A$_2$=Na, K$_2$ and NaK) were synthesized by a solid-state reaction process. High-purity CuO, CuSO$_4$, and A$_2$SO$_4$ (A=Na, K) were mixed in a molar ratio 1:2:1, followed by annealing at $500^\circ$~C,  $580^\circ$~C and $480^\circ$~C in air (for Na, K and NaK, respectively) for at least 100 h with intermediate grindings. The samples were characterized by X-ray and neutron diffraction, confirming their single-phase character.

Neutron scattering measurements for the magnetic structure determination were performed at the time-of-flight Cold Neutron Chopper Spectrometer (CNCS)~\cite{CNCS} at the Spallation Neutron Source at Oak Ridge National Laboratory (USA). The samples were enclosed in aluminum cylinders (8-mm diameter, 45-mm height) and placed into an orange cryostat. The data were collected using fixed incident neutron wavelength $\lambda = 4.96$~\AA.
For the crystal structure determination, we used the high-resolution diffractometer for thermal neutrons HRPT \cite{hrpt} at the SINQ spallation source at the Paul Scherrer Institute (Switzerland) using wavelength  2.45~\AA.  

The determination of the crystal and magnetic structure parameters were done using the {\tt FULLPROF}~\cite{Fullprof} program, with the use of its internal tables for neutron scattering lengths. The symmetry analysis was performed using {\tt ISODISTORT} from the {\tt ISOTROPY} software \cite{isod,isod2} and some software tools of the Bilbao crystallographic server such as  {\tt MVISUALIZE} \cite{Bilbao,ISI:000358484200010}.

\section{Crystal structure}
\label{cryst_sec}

%%
%Text according to your draft. You may reference the first structure studies, see references 3-5 in our publication Phys. Rev. B 98, 180410(R) (2018).
%% 

The crystal structures in all samples \acuos\ (A$_2$=NaK, K$_2$, Na$_2$) are well refined in the monoclinic space group $C 2/c$ (no. 15) from the HRPT data with the example of the diffraction pattern and its Rietveld refinement shown in Fig.~\ref{dif_HRPT}.  The structure parameters are listed in the supplementary materials in Tables S1 and S2 in Refs.~\cite{Furrer2018} and~\cite{Furrer2021}, respectively. Here we present in Table ~\ref{tab_Cu} only the crystal metrics and the Cu positions relevant for the present paper. The full structure information  can be found  in magnetic crystallographic information files (mcif) in the Supplemental Material~\cite{SM}.
%%%%% TODO: We have to give a table with full structure parameters in SM. We will need it anyway for the mcif-files. So they have to be in both settings - original C2/c and final in MSG if there is a basis transformation.

One unit cell contains 24 Cu atoms forming four Cu-hexamers, two of which are shown in Fig.~\ref{hexamer}. Each hexamer has 2-fold axis symmetry and formed by Cu1, Cu2 and Cu3 pairs. The atoms in the pairs are related by a 180~deg rotation around the y-axis ($\{2_y|00{1\over2}\}$ operator in $C 2/c$). The second hexamer is related to the first one by two other symmetry operators -1 and $\{m_y|00{1\over2}\}$.

The diffraction patterns measured at CNCS with large neutron wavelength are limited by the Q-range ($Q_{\rm max}$ = 2.3~\AA$^{-1}$) so that in the crystal structure fits only the profile parameters and overall scale factor were refined to be used later for the magnetic structure fits. The fit quality is good for all three samples as illustrated for A$_2$=NaK in Fig.~\ref{dif_CNCS}. 

\section{Expectation values of $1\over2$-spins in an isolated hexamer}
\label{triplet}

The Hamiltonian of the isolated hexamer reads:

\beq{H}
H = -2 \sum_{i=1}^7 J_{i} \vec{s} \cdot \vec{s}{'} - h_{mf} S_z
\label{Hamiltonian}
\eeq

where the sum runs over the exchange couplings $J_i$ along the bonds in the hexamer shown in Fig.~\ref{hexamer}, $h_{mf}$ is the mean field and $S_z$ is the total spin projection of the hexamer molecule. Based on the results of inelastic neutron scattering INS studies~\cite{Furrer2018,Furrer2020,Furrer2021} we have two models.
% In general one can have 15 independent exchange parameters, but we limit the consideration up to seven parameters according to the INS studies~\cite{Furrer2020}. 
%
The symmetric model with 4 parameters is solved exactly. In this case the Hamiltonian has different $J_1$, $J_3$, $J_2$ between Cu1, Cu2 and Cu3 spin pairs and the same cross exchange couplings $J_5$, $J_6$, $J_7$ equal to $J_4$. 
The four parameters symmetric Hamiltonian is diagonalized algebraically with explicit formula for the energy given by Formula (2) in the Ref.~\cite{Furrer2020} in the coupled basis with 5 quantum numbers, being sums of spins in pairs Cu1, Cu2, Cu3, the sum of the pairs Cu1 and Cu2 and the total spin S. 

The individual spins do not commute with the Hamiltonian, but the spin expectations values \av{s} are derived straightforwardly for the above pure quantum states through the Clebsh-Gordon coefficients and do not depend on the exchange constants $J_i$. Only the energy spectrum depends on the exchange parameters $J_i$.
Here we use the basis of individual spins in the hexamer to calculate the expectation values of the spins to be later compared with the experiment for the general model of the Hamiltonian from INS-experiments. In the spectrum, we have five singlets, nine triplets, five quintets, and one septet with the rational spin expectation values \av{s} of the Cu spins. 
% The wave functions of all multiplets and the spin expectations values of the spins \av{s} do not depend on the specific exchange values of $J_i$. 

%
For the experimentally determined exchange parameters $J_i$ ($i=1..4$) 1.2, 12.5, 2.3, and -6.7~meV  for sample with A=Na~\cite{Furrer2020}, the ground state is a triplet and the next exited state is a singlet separated by the energy gap $2 J_4 = 13.4$~meV. The mean field on the hexamer lifting up the degeneracy is less than $h_{mf}<1.5$~meV~\cite{Furrer2021} and does not mix the triplet wave function with the next singlet state. 
The exchange parameters are slightly different for K and NaK samples, but the ground state is the same triplet with spin expectation values $\av{s}={3\over8}$ %(${3\over4}$~$\mu{\rm B}$) 
for the side Cu1 and Cu2, and $\av{s}=-{1\over4}$ 
%($-{1\over2}$~$\mu{\rm B}$) 
for Cu3 in the middle for all three compositions. The spins in the Cu1, Cu2 and Cu3 pairs are coupled ferromagnetically (FM), and antiferromagnetically (AFM) between the middle Cu3 pair and the side Cu1 and Cu2 pairs with total hexamer spin $S_z=1$. As we will show further in Sec.~\ref{mag_det} this configuration is indeed observed experimentally with magnetic moment values in reasonable agreement with the above theoretical values.
%
%Interestingly, for this low spin 1/2 hexamer the antiferromagnetic (AM) configuration inside the hexamer is possible only for the triplets ...
%
In the model with seven independent exchange parameters $J_i$ ($i=1..7$): 1.3, 11.5, 2.2, -8.3, -8.3, -4.7, -5.3 refined from INS experiment~\cite{Furrer2021} the Cu spins are coupled the same way. The Cu1, Cu3 and Cu2 spin expectation values \av{s} amounted to 0.395, -0.246, 0.351  similar to the symmetric model. The total hexamer spin projection remains a good quantum number $S_z=1$. The magnetic moment expectation values with the seven parameters from Ref.~\cite{Furrer2021} are the same within the accuracy 0.005~$\mu \rm_{B}$  for all three samples, being 2\av{s} = 0.79, 0.49, 0.70~$\mu \rm_{B}$  for Cu1, Cu3 and Cu2 magnetic moments. Actually the FM coupling in the side pairs is not very sensitive to the small exchange parameters $J_1$ and $J_3$ in comparison with the large FM exchange $J_2$ in the middle Cu3-pair and inter-pair AFM exchanges $J_i$ ($i=4..7$). The ground state will be the same triplet with FM-pairs even if one flips the sign of the $J_1$ and $J_3$.
%
%symmetric: Vector[row](6, [0.375, 0.375, -0.250, -0.250, 0.375, 0.375])
%
%Na:  1.3, 11.5, 2.2, -8.3, -8.3, -4.7, -5.3
%
%general: Vector[row](6, [0.395, 0.395, -0.246, -0.246, 0.351, 0.351])
%Vector[row](6, [0.790, 0.790, -0.492, -0.492, 0.702, 0.702])
%
%NaK: 1.8, 12.3, 2.9, -8.1, -8.1, -4.6, -5.1
%
%Vector[row](6, [0.396, 0.396, -0.247, -0.247, 0.351, 0.351])
%Vector[row](6, [0.792, 0.792, -0.494, -0.494, 0.702, 0.702])
%
%K2 : 1.6, 12.4, 2.5, -7.9, -7.9, -4.6, -5.2 
%
%Vector[row](6, [0.395, 0.395, -0.247, -0.247, 0.352, 0.352])
%Vector[row](6, [0.790, 0.790, -0.494, -0.494, 0.704, 0.704])

\section{Magnetic structures}
\label{mag_str}

The neutron diffraction intensities are dominated by very large nuclear peaks. One can see only subtle differences between base and paramagnetic temperatures on the raw neutron diffraction patterns as shown in Fig. \ref{dif_raw}. The magnetic scattering is small and amounts to only less than about 0.5\% in intensity as one can see from the figure.
For this reason, the difference patterns, i.e. the difference between patterns taken at base ($\simeq1.6$~K) and paramagnetic (6~K) temperatures, were used to solve and refine the magnetic structures. Such difference patterns contain purely magnetic scattering and are free of possible systematic uncertainties due to the fitting  of large crystal structure Bragg peaks, background, impurities, etc.

The identification of the magnetic propagation vector was done using the so called le~Bail fitting, where all peak intensities are refined separately without any structure model, thus allowing a straightforward determination of the propagation vector $\bm k$. In addition this model free fit defines the best possible goodness of fit. We have found that for the samples with A$_2$ =K$_2$ and NaK the propagation vector is ${\bm k}$=0 or gamma point (GM) of Brillouin zone (BZ) [here we use internationally established nomenclature for the irreducible representations (irreps) labels  and magnetic superspace groups MSSG~\cite{Bilbao,isod}]. The fit quality for the sample with A=K is illustrated in Fig.~\ref{dif_leBail_K}. The total number of magnetic Bragg peaks is about 120. For the sample with A=Na the propagation vector is $\bm k$=[0,1,0] (Y-point of BZ). The goodness of Le Bail fits is shown in Table~\ref{magmom}.
% /Users/pomjakushin/work/Fits/Albert/K2Cu3O(SO4)3/hrpt/fin/NaK/andrey/simann/NaK_1p7K-6K_mon1000_add50_extract_lebail_profile.sum
% Rp:  6.06     Rwp:  6.91     Rexp:    7.03 Chi2: 0.966 
% /Users/pomjakushin/work/Fits/Albert/K2Cu3O(SO4)3/hrpt/fin/K/Andrey_K/simann/K2_1p6K-6K_mon1000_add50_extract_lebail_profile.sum
% Rp:  6.15     Rwp:  5.88     Rexp:    4.65 Chi2:  1.60
% /Users/pomjakushin/work/Fits/Albert/K2Cu3O(SO4)3/hrpt/fin/Na/Andrey/simann/Na_1p7K-6K_mon1000_add50_extract_lebail_profile.sum
% Rp:  5.65     Rwp:  6.36     Rexp:    5.91 Chi2:  1.16 GM
%/Users/pomjakushin/work/Fits/Albert/K2Cu3O(SO4)3/hrpt/fin/Na/Andrey/Na2_1p5K-6K_mon1000_add50_ksearch.sum
% 6.00     Rwp:  6.29     Rexp:    5.32 Chi2:  1.40   Y

\subsection{Symmetry analysis}
\label{symmetry}

The parent space group C2/c (no. 15) has four irreps for the gamma point GM $[0,0,0]$ of the BZ. Since they are all one-dimensional real irreps, there are only four Shubnikov magnetic space subgroups (MSG): mGM1+ C2/c 15.85, mGM2+ 15.89 C2'/c', mGM1- 15.88 C2/c', mGM2- 15.87 C2'/c (we use unified UNI MSG symbol~\cite{isod},\cite{Campbell_UNI}). The first two, which are inversion even (i.e. -1 is not primed), allow ferromagnetic (FM) ordering. For each MSG there are nine independent parameters to be determined experimentally. For the Y-point $[0,1,0]$ there are four irreps as well that generate four magnetic models in the MSGs 14.84 P2\_1/c.1'\_C[C2/c] and 13.74 P2/c.1'\_C[C2/c] with two different basis transformations from the parent space group.  

%_space_group_magn.number_BNS "14.84"
%_space_group_magn.name_UNI "P2_1/c.1'_C[C2/c]"
%_space_group_magn.name_BNS "P_C2_1/c"
%_space_group_magn.number_OG "15.7.98"
%_space_group_magn.name_OG "C_P2'/c"
%
%_space_group_magn.number_BNS "13.74"
%_space_group_magn.name_UNI "P2/c.1'_C[C2/c]"
%_space_group_magn.name_BNS "P_C2/c"
%_space_group_magn.number_OG "15.6.97"
%_space_group_magn.name_OG "C_P2/c"
%_space_group_magn.point_group_number "5.2.13"
%_space_group_magn.point_group_name_UNI "2/m.1'"
% 
We present some more details for the MSG C2'/c (mGM2-), which is the solution that fit the neutron diffraction data very well for the samples with A$_2$=K$_2$, NaK  and MSG P2\_1/c.1'\_C[C2/c] (mY2-) for the sample with A=Na, as shown in the next section \ref{mag_det}.
In both cases there are 4 hexamers per unit cell (Fig.~\ref{hexamer},\ref{mag_strs}), but the magnetic structure is defined by the magnetic configuration in a single hexamer. 

The basis transformation to the C2'/c MSG structure is identity matrix with the zero origin shift. The hexamers are FM coupled in the ab-plane because they are related by C-centering, which does not change the spin for the GM-point. The coupling between closest hexamers along ($bc$)-diagonal related by primed inversion -1' is AFM. The above inter-hexamer couplings are fixed by the magnetic symmetry. The couplings inside the hexamers are only partly fixed by symmetry. The couplings between different copper spin pairs formed by Cu1, Cu2 and Cu3 positions are not fixed by symmetry and should be found from the fits to the experimental data. The coupling inside each Cu-pair is FM in the $ac$-plane and AFM along $b$-axis because the spins in the pair are related by primed 2-fold axis $2'_y$. One can go down to lower symmetry by mixing the irreps mGM2- and mGM2+, resulting in the subgroup C2'. In this group the intra-hexamer coupling remains the same, but the second hexamer becomes independent, doubling the number of free parameters to 18. If one adds additionally irrep mGM1+, then the symmetry relations between all 24 atoms in both hexamers disappear in the MSG P1. 

The basis transformation to the P2\_1/c.1'\_C[C2/c] 14.84 (mY2-) MSG structure is identity matrix with the origin shift (1/4,1/4,0). Similarly to the C2'/c MSG, the hexamers are completely related by symmetry. Technically, by symmetry operators, the pairs in the hexamers are constructed differently. The inversion -1 relates the spins on different hexamers as before, but the two-fold screw axis $2_1$, now relates the spins in the different hexamers as well. The second spin in each pair is generated by the time-odd centering translation 1'\_C operator.

Although the accepted standard symbol and settings of the MSG is UNI, which combines a modified BNS symbol with essential information from the OG symbol~\cite{isod},\cite{Campbell_UNI}, we like to give also the description of this MSG in OG-setting, which is, in our view, more physically relevant here. In this setting there is no origin shift from the parent group and the MSG OG-symbol reads C\_P2'/c. So the Cu-spins in the Cu-pairs generated by non-centered operators are have exactly the same configuration as for C2'/c, but the pairs that are related by C-centering are time reversal odd due to OG propagation vector $\bm k$=[0,1,0]. Finally the only difference of the magnetic structures in OG-settings is that the hexamers that are related by centering translations (1/2,1/2,0)+ are coupled AFM in sample with A=Na, but not FM as in the samples with A$_2$ =K$_2$ and NaK  (Fig.~\ref{mag_strs}).

Interestingly, there is another different magnetic structure in the same MSG (no.14.84) but generated by the parity even irrep mY2+. The structure is derived from the parent grey space group by different basis transformation and in spite of the same symmetry elements the magnetic structure is different. This emphasizes the importance of specifying the new atomic positions in the MSG if the basis transformation is not given. %This second solution in the same MSG does not fit the experimental data.

%%%%%%%%%%%
% the business with 1/4,1/4,1/4 shift should be clarified
% The structure table should have explicit abc and Cu positions, 
% In the magtable we need to have MSG in the caption
% I have to compare the xyz, abc from cif and mcif! And also create the mcif files
% the basis transform from parent should be specified.
%%%%%%%%%%%
% The less symmetric group P21/c with 4 symops is augmented with (1/2,1/2,0) centering translation and becomes C2/c. The origin shift (3/4,3/4,0) converts 21=>2,  c=>c; -1 => -1 (1/2,1/2,0)+, see previous slide.

\subsection{Magnetic structure determination}
\label{mag_det}

As a first step, using the {\tt FULLPROF} program, we have performed a simulated annealing (SA) search \cite{kirkpatrick83,Fullprof} of the full diffraction profile for the models described in the previous sections.  A SA search  starts from random values of the free parameters and we have repeated the search more than several hundreds times. The search was performed for all three samples. The results of SA-search are listed in table \ref{magmom} and the  illustrations of the magnetic structures corresponding to SA part in the table are shown in Fig.~\ref{mag_strs}. Note that the parameters found in  SA search do not have error bars.

%NB: the distances between hexamers related by -1 is the shortest about 7~\AA, whereas the distance between ones related by C-centering is larger about 10~\AA. This might explain the 1'\_C in Na-sample.

Based on the values from SA-search we have performed standard Rielveld least square (LSQ) refinement. Some parameters had to be correlated or fixed to get the convergent fit. 
Figures \ref{dif_mag_NaK}, \ref{dif_mag_K} and \ref{dif_mag_Na} show the LSQ-fits of the difference diffraction patterns. Although the magnetic structures that we have found are similar for the samples with A$_2$ =K$_2$ and NaK, the diffraction patterns have different distribution of the intensities over the Bragg peaks  For instance the first Bragg peak (001) at $2\theta\simeq20$~deg  has different relative intensity in comparison with the main group of peaks in the middle of the pattern around $2\theta\simeq70$~deg. The sample with A=Na has completely different diffraction pattern, e.g. the first peak has zero intensity. 
For the sample with A=K we present two types of fits in the Table~\ref{magmom}: (i) the middle spin Cu3 angles are fixed to $\theta=90$~deg and $\phi=90$~deg, leaving only z-component of the Cu3-spin to be non-zero, and (ii) the spherical angles and moment sizes are constrained to be the same for Cu1 and Cu2 spins, making them ideally parallel. %This is not needed if one refines the background... This is NOT true.
For the for the sample with A$_2$=NaK the LSQ-fit converges with all parameters released, but for completeness we present also the constrained fit with the spherical angles to be the same for Cu1 and Cu2 spins, similar to  the sample with A=K. For the sample with A=Na we present two types of fits similar to the sample with A=K.

% decrease in intensity between Q=1.4 and 2.2A-1 due to magnetic form-factoris about 40\%, so small/large intensities are mainly by the magmodel.

%%The problem with Na-fit is not only absence of first (001) peak but also  the praks arounf 80 degrees. So it is not the additional scattering...
%%
%%For the K-sample SA finds two equivalent solutions related by time inversion 1' with the parameters listed in Table~\ref{magmom}. 
%% 
%% The reliability profile factors $Rp$ (in percents) for the solutions came in the ranges 2.44~-~2.5, 2.46 - 2.56 and 2.45 - 2.46 for the 3+1, 3+3 and 3+2 magnetic models, respectively.
%%We have 3*2 angles (phi+theta), and the same absolute spin size.
%%Then SA for all?. 
%%
%%Models to consider:
%%1. all free in A 1 1 2/a: 6 parameters
%%
%%K-one: Rp:  6.45     Rwp:  6.04     Rexp:    4.65 Chi2:  1.69 

\section{Discussion}
\label{discon}

We have synthesized and studied the crystal and magnetic structures of the compounds \acuos\  (A$_2$=Na$_2$, NaK, K$_2$). Crystallographically, the $S$=1 copper hexamers form a two-dimensional network in the ($bc$)-plane, but three-dimensional long-range magnetic order is observed below $T_N$. Surprisingly, there are differences in the long-range magnetic order: the compounds with A$_2$=NaK and A=K are defined by the magnetic Shubnikov group C2'/c, whereas the compound with A=Na is governed by P2\_1/c.1'\_C[C2/c] in standard UNI-notation, as explained in section~\ref{symmetry}. Here for the comparison of the two structures we prefer to use Opechowski-Guccione (OG) notation~\cite{MSG}. In this notation both groups C2'/c and C\_P2'/c keep the parent space group metrics and have the same set of symmetry operators for non-centered atoms, but the atoms related by centering translations (1/2,1/2,0)+ are FM coupled in C2'/c and AFM in C\_P2'/c.  In both groups the copper spins in Cu-pairs are related by symmetry and coupled FM for spin components in the $(ac)$-plane and AFM along $b$-axis. Figure~\ref{mag_strs} shows the magnetic structures for all three compounds with the simulated annealing values from Table~\ref{magmom}. As one can see from the Table~\ref{magmom} and Fig.~\ref{mag_strs}, the largest spin component is along z-axis that makes the Cu-pairs approximately FM aligned. The neighboring pairs in hexamers are approximately AFM coupled, and this coupling is not dictated by symmetry, but is a result of the fits of the experimental data. Figure~\ref{K_mag_projections} further illustrates the spin directions in three projections for K-sample. The ($ac$)- and ($bc$)-projections nicely demonstrate the couplings and magnetic moment directions. One can also see that the centered hexamers are FM-coupled in the ($ab$)-projection.

%Maybe give the calculated values in the table... with g-factor =2.
%:
The ground state is expected to be the triplet with FM-coupling in the Cu-pairs and AFM between neighboring Cu pairs as was shown in section \ref{triplet}. As  discussed above the main spin components are in accordance with this type of coupling. 
However the presence of the noncollinear components of the magnetic moments within the Cu hexamers gives evidence for spin frustration. We think that the origin of the noncollinear spin arrangement results from a compromise to handle possible spin frustrations within the weak interhexamer interactions.
From the isolated hexamer calculations we have the magnetic moments 0.79, 0.70, and 0.49~$\mu \rm_{B}$  (using g-factor 2) for the end positions Cu1, Cu2 and middle Cu3. The sizes of the magnetic moments are expectedly smaller than the Cu$^{2+}$ ion value of 1~$\mu \rm_{B}$  due to quantum entanglement in the wave function in hexamer. Experimentally we have for Cu1, Cu3 and Cu2 the following values of magnetic moment sizes 
K:  0.59(3),  0.44(2),  0.43(4), NaK: 0.62(5), 0.53(5), 0.58(8), Na: 0.73(4), 0.48(2), 0.49(5) in $\mu \rm_{B}$. We find that this is a fairly good correspondence between the experiment and the calculations in the simple model disregarding the inter-hexamer interaction. In particularly, for the middle pair Cu3 the calculated moment 0.49~$\mu \rm_{B}$  is practically within the experimental error bars. The side Cu1 and Cu2 pairs have experimentally smaller values (by maximum 26\% and 39\% smaller in the sample with A=K) than expected from the isolated hexamer. This is not surprising because the strongest inter-hexamer interactions that can modify the spin expectation values, are between the end-standing Cu spins as discussed below. 

Nekrasova et al. performed ab initio calculations to estimate the intrahexamer and interhexamer superexchange interactions and the anisotropy for the hexamer molecule~\cite{Tsirlin}. The strongest calculated intrahexamer interactions of type Cu-O-Cu are in amazingly good agreement with the results derived directly from neutron spectroscopy INS experiments \cite{Furrer2018,Furrer2020,Furrer2021}. The two smallest interactions in the end-standing Cu1-Cu1 and Cu2-Cu2 pairs is calculated to be antiferromagnetic in contradiction with the INS experiments, where these two couplings were ferromagnetic
~\footnote{The signs of $J_{\rm neutron}$ in Ref.~\cite{Tsirlin} (Table II) for the end pairs for A=K are reversed in comparison with original paper~\cite{Furrer2020}. The sizes of $J_{\rm neutron}$ are correct. The numeration of Cu in the hexamer is different from the one in present paper.}. 
%
%This discrepancy indicates that ab initio calculations have to be considered with caution, even for the case of simple Cu-O-Cu superexchange bridges.
%
In principle, the FM sign of the intrahexamer interactions of type Cu-O-Cu in Cu-pairs is consistent with the Goodenough-Kanamori-Anderson rules.
Unfortunately the above ab initio calculations did not show the spin expectation values, that could be then compared with the experimental values from the present diffraction data and with our calculations for the isolated hexamers in the mean field. 

\begin{figure}
  \begin{center}
    \includegraphics[width=\count1cm]{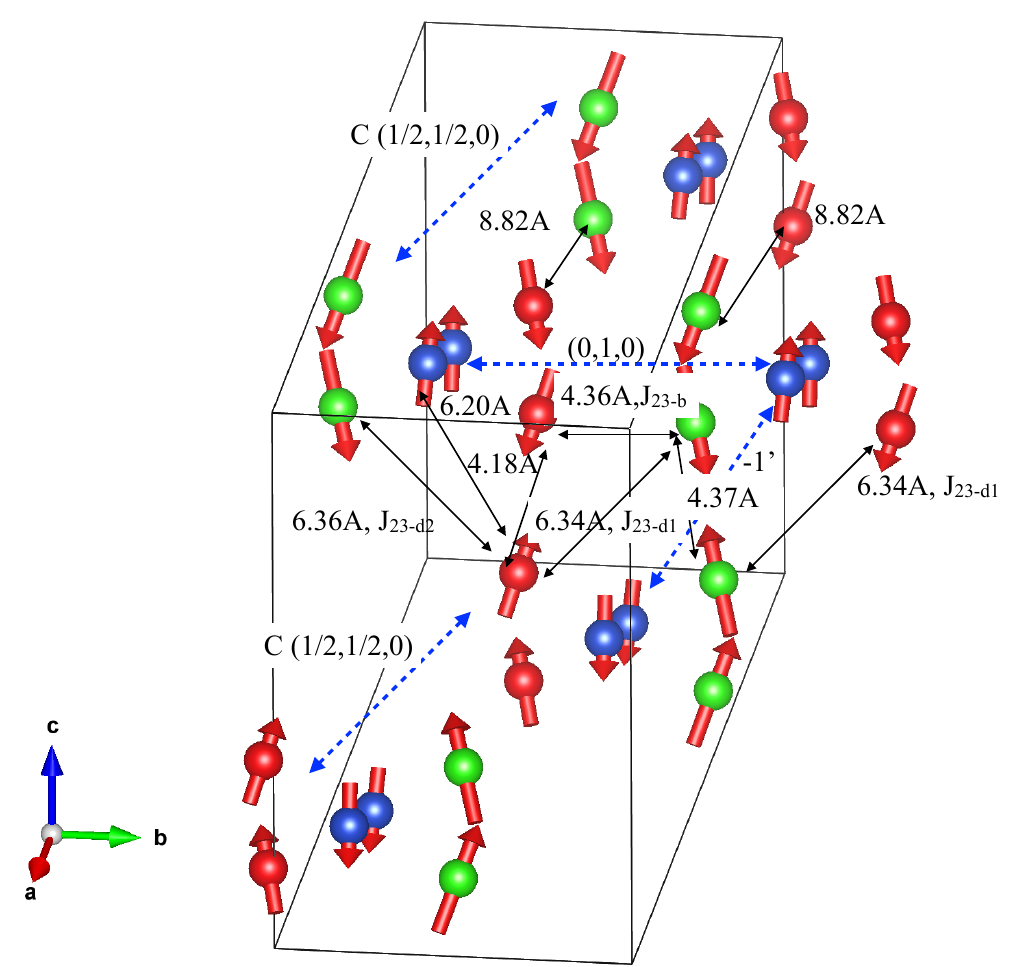}
 \end{center}
\caption{Magnetic structure in the sample with A=K (C2'/c) showing the connectivity between hexamers. Side spin pairs Cu1 and Cu2 are in green and red, the central Cu3 pair is in blue. Black arrows indicate bonds between side hexamer spins Cu1 and Cu2 between different hexamers. Three bonds are labeled in the notation from Ref.~\cite{Tsirlin}. Five hexamers are shown with the transforming symmetry operations indicated by blue dashed lines. The hexamers that are AFM coupled are related by inversion -1'. The ones that FM coupled are related by C-centering (1/2,1/2,0) or just translation (0,1,0)}
\label{mag_5hex}
\end{figure}

Figure~\ref{mag_5hex} shows five hexamers in the magnetic structure in the K-sample with several bonds between Cu atoms indicated by black arrows. The blue dashed arrows show the symmetry relations between the hexamers. One hexamer from the neighboring cell shifted by (0,1,0) is shown to better visualize the bonding. Figure~\ref{K_mag_projections} further illustrates the inter-hexamer coupling in three crystallographic projection. The hexamers along $a$, $b$ and $c$ directions are always FM-arranged because they are separated by integer lattice translations and the k-vector is zero or one. These distances between Cu spins are very large about 19 and 14~\AA\ for $a$ and $b$ directions. However, since the hexamer is elongated mainly along $b$-direction the inter-distances between Cu1 and Cu2 from neighboring hexamers along $b$-axis are small about 4.36~\AA, as shown in the Figure~\ref{mag_5hex} and labeled as $J_{23-b}$ following the notation~\cite{Tsirlin}. The hexamers along the $ab$, $bc$ and $ac$ diagonals belong to the same unit cell and their coupling depends on the specific magnetic symmetry.
The ab initio calculations give zero coupling along the b-axis \cite{Tsirlin}, which is in contrast to the ferromagnetic couplings obtained from the analysis of the magnetic excitation spectra \cite{Furrer2021}. Experimentally we have FM arrangement in this direction as explained above.
The ab initio calculations predict the strongest AFM couplings along the diagonals in the ($bc$)-plane as indicated in Fig.~\ref{mag_5hex} by $J_{23-d1}$ and $J_{23-d2}$ bonds. This diagonal $bc$-coupling is indeed AFM as observed experimentally for all three compounds, and these are the bonds between Cu1 and Cu2 atoms in the hexamers belonging to the same unit cell and related by inversion. The AFM coupling along $bc$ diagonal is also well seen in Fig.~\ref{K_mag_projections}. The hexamer spins arrangement along $ab$ diagonal is given by C-centering translation, and it is FM in the samples with A$_2$=K$_2$, NaK as one can see in Fig.~\ref{K_mag_projections} and AFM for the Na-sample. As a consequence the arrangement along $ac$ diagonal is AFM in K$_2$ and NaK samples and FM in the Na-sample.  In general, all inter-hexamer coupling in all three samples are protected by magnetic symmetry. The difference in the magnetic ordering in the Na-sample is apparently due to different coupling along $ab$ and $ac$ diagonals. These couplings are independent from the coupling along $bc$ diagonal. It would be interesting to compare with the ab initio theoretical calculations to see the reason for the different coupling sign.
The fact that the lattice parameter ratios $a/b$ and $a/c$ for A=Na differ by almost 10\% from those for A=K and A$_2$=NaK may be indicative of the interhexamer couplings in A=Na being different from those in  A$_2$=K and NaK.

\section{Conclusions}
\label{conclusion}

The crystal and magnetic structures of the spin $S$=1 hexamer cluster  $\acuos$ (fedotovite A$_2$=K$_2$, euchlorine NaK, puninite Na$_2$) were studied by neutron powder diffraction at temperatures 1.6-290 K. The crystal structures in all compounds are well refined in the monoclinic space group $C 2/c$ (no 15), whereas the magnetic structures are different with the magnetic space groups MSGs 15.87 C2'/c for A$_2$=K$_2$, NaK and 14.84 P2\_1/c.1'\_C[C2/c] for A=Na.
The basic magnetic units of the compounds are strongly coupled pairs of Cu in quasi-isolated copper hexamers which are coupled to each other by weak superexchange interactions. The inter-hexamer interactions are completely fixed by MSGs. The coupling of the Cu spins in pairs is fixed by MSG symmetry, whereas the inter-pair coupling is not symmetry related. Experimentally we found that the pairs are approximately ferromagnetically FM coupled in pairs and antiferromagnetically AFM between neighboring pairs. The magnetic moment sizes for side pairs Cu1, Cu2 and middle Cu3 spins amounted to K:  0.59(3), 0.43(4), 0.44(2), NaK: 0.62(5),  0.58(8), 0.53(5), and Na: 0.73(4),  0.49(5), 0.48(2) in $\mu \rm_{B}$. The coupling and the moment sizes are in good accordance with the model calculations for the isolated hexamers in the mean field with Hamiltonian parameters from inelastic neutron scattering INS studies~\cite{Furrer2018,Furrer2020,Furrer2021}. The ground state is a triplet for the Hamiltonian with FM in pairs and AFM inter-pair coupling with the spin expectation values \av{s} independent on the exchange constants being 2\av{s}=3/4~$\mu \rm_{B}$  for the end pairs and  1/2~$\mu \rm_{B}$  for the middle pair. For more refined exchange parameters the 2\av{s} are the same within the accuracy 0.005~$\mu \rm_{B}$  for all three samples, being 0.79, 0.70, 0.49~$\mu \rm_{B}$  for the side pairs Cu1, Cu2 and middle pair Cu3 magnetic moments.

\section*{A{\lowercase{cknowledgements}}}

Part of this work was performed at the Swiss Spallation Neutron Source (SINQ), Paul Scherrer Institut (PSI), Villigen, Switzerland. This research used resources at the Spallation Neutron Source, a DOE Office of Science User Facility operated by the Oak Ridge National Laboratory.
V.P. thanks V.~Markushin for the discussions.

\bibliography{Mag_str_A2Cu3O_SO4_3_arxiv_PRB.bbl}

%===========================      ========================
%                           Tables
%===========================      ========================
%
%  In case if I decide to give the tables on crystal structure in this paper
%
% [pomjakusX:~] pomjakushin% cd '/Users/pomjakushin/work/Fits/Albert/K2Cu3O(SO4)3/hrpt/fin'
% [pomjakusX:K2Cu3O(SO4)3/hrpt/fin] pomjakushin% ./extract.pl 
%

\newpage
\setlength{\LTcapwidth}{\textwidth} \setlength\LTleft{-10pt}
\setlength\LTright{0pt} \setlongtables
\begin{longtable}{l|p{0.3\textwidth}|p{0.3\textwidth}|p{0.3\textwidth}}

\caption{The structure parameters in \acuos\ (A$_2$=K$_2$, NaK, Na$_2$)  at T=2K in the parent space group C2/c: lattice parameters $a, b, c,$ $\beta$ and fractional atomic coordinates $x, y, z$. The positions of Cu are the same in the magnetic space group MSG 15.87 C2'/c for A$_2$=K$_2$, NaK.   For A=Na the MSG is 14.84 P2\_1/c.1'\_C[C2/c], the Cu positions are different due to origin shift and given on separate line. All atoms are in the Wyckoff positions (8f). Only Cu atoms coordinates are presented that are relevant for the analysis of the magnetic diffraction patterns. Conventional reliability factors~\cite{Fullprof}  $R_{p},$ $R_{wp}$, $R_{exp}$, $\chi^2$ are also given. The parameters were refined from the powder neutron diffraction patterns measured at HRPT/SINQ with wavelength $\lambda=2.45$~\AA. }\\

\label{tab_Cu}

%\begin{center}
%\begin{tabular}{l|[6cm]p|p|p}

                           &A$_2$=K$_2$                               &A$_2$=NaK                               &A$_2$=Na$_2$ \\
\hline
$a$, \AA            & 18.97550(66)                      & 18.47297(75)                & 17.21406(67)                        \\
$b$, \AA            & 9.50038(35)                       &  9.36442(38)                &  9.37286(35)                        \\
$c$, \AA            & 14.19721(51)                      & 14.31463(57)                & 14.37014(54)                        \\
$\gamma$, deg       & 110.49150(85)                     & 113.96423(98)               & 111.84364(75)                        \\
$V$, $\AA^3$        & 2397.45(15)                       & 2262.81(16)                 & 2152.08(14)                           \\
\hline
Cu1 $x\, y\, z$         &  0.48127(27)  0.02265(54)  0.34025(32) & 0.48267(28)  0.02045(55)  0.34304(37)  & 0.47740(24)  0.02021(53)  0.34063(31) \\
Cu2 $x\, y\, z$         &  0.48577(23)  0.47775(52)  0.13924(33) & 0.48341(25)  0.47239(51)  0.13814(35)  & 0.48559(23)  0.47886(44)  0.14103(31) \\
Cu3 $x\, y\, z$         &  0.41989(28)  0.74506(65)  0.20554(35) & 0.41661(27)  0.74829(73)  0.20431(34)  & 0.41231(25)  0.74746(49)  0.20169(33) \\
\hline
R-factors & 3.58  4.65    3.00  2.41 &  3.42  4.39    3.16  1.93 & 3.39  4.45    2.99  2.20 \\
%\multicolumn{2}{l|}{$R_{p}, R_{wp}, R_{exp}, \chi^2$}&
\hline
\multicolumn{4}{c}{A=Na, Cu positions in MSG P2\_1/c.1'\_C[C2/c]}\\
\hline
                           &Cu1 $x\, y\, z$               &Cu2 $x\, y\, z$       &Cu3 $x\, y\, z$\\
               & 0.22740  0.77021  0.34063 & 0.23559  0.22886  0.14103 & 0.66231 -0.00250  0.20169 \\

% Cu1 $x,y,z$         & 0.48127, 0.02265, 0.34025 &0.48127, 0.02265, 0.34025 &0.48127, 0.02265, 0.34025 \\ % this will fit...
\hline

%\end{tabular}
%\end{center}
\end{longtable}
\newpage

%%%%%%%%%%%%%%%%%%%%%%%%%%%%%%%%%%%%%%%%%%%%%%%%%%%
%
%  Magnetic structures 
%
%%%%%%%%%%%%%%%%%%%%%%%%%%%%%%%%%%%%%%%%%%%%%%%%%%%
\newpage
%%%% This is for longtable
\setlength{\LTcapwidth}{\textwidth} \setlength\LTleft{0pt}
\setlength\LTright{0pt} \setlongtables

\begin{longtable}{l|l|l|l|l|l|l|l|l|l} % 10 columns
%%%% --------------------

%%%% This is for table
%\begin{table}
%%%% --------------------

\caption{Magnetic model parameters for \acuos\ (A$_2$=K$_2$, NaK, Na$_2$) refined from
the   diffraction data shown in Figures~\ref{dif_mag_NaK},\ref{dif_mag_K},\ref{dif_mag_Na}. The numeration of the
  atoms is as indicated in Table~\ref{tab_Cu}. $M$ is the size of the
  magnetic moment, $\phi$ and $\theta$ are spherical angles with $c$
  and $b$ axes, respectively. $m_x$, $m_y$ and $m_z$ are respective magnetic moment projections in the MSGs C2'/c for A$_2$=K$_2$, NaK and P2\_1/c.1'\_C[C2/c] for A=Na. The error bars are given only for the refined parameters. The reliability factors~\cite{Fullprof} $R_{p}, R_{wp}, R_{exp}, \chi^2$ are shown as the bottom lines for each type of fit. The magnetic crystallographic information (mcif) files for LSQ-fits can be found in the Supplemental Material~\cite{SM}.
  }
\label{magmom}\\ % these \\ are needed! for longtable

%\begin{center}
%%%% This is for table
%\begin{tabular}{l|l|l|l|l|l|l|l|l|l}
%%%% --------------------

     &  \multicolumn{3}{c|}{K} & \multicolumn{3}{c|}{NaK} &  \multicolumn{3}{c}{Na}  \\\hline
     &$M, \mu_B$       & $\phi$     & $\theta$ 
     &$M, \mu_B$       & $\phi$     & $\theta$ 
     &$M, \mu_B$       & $\phi$     & $\theta$ \\ 

     &$m_x$       & $m_y$     &  $m_z$ 
     &$m_x$       & $m_y$     &  $m_z$ 
     &$m_x$       & $m_y$     &  $m_z$ \\ \hline

\multicolumn{10}{l}{ Le Bail model free fit}\\ \hline     

\multicolumn{2}{l|}{$R_{p}, R_{wp}, R_{exp}, \chi^2$}&
\multicolumn{2}{l|}{6.15, 5.88, 4.65, 1.60}  & 
\multicolumn{3}{l|}{6.06,  6.91,    7.03, 0.966 } & 
% \multicolumn{3}{l}{5.65,  6.36,    5.91,  1.16 } \\\hline
% \multicolumn{3}{l}{5.97,  6.94,    6.12,  1.29  } \\\hline % profile_B Na
\multicolumn{3}{l}{ 6.00,  6.29,    5.32,  1.40    } \\\hline % Y-point

\multicolumn{10}{l}{ Simulated annealing (SA)}\\ \hline
%% Editing table BEGIN
Cu1   & 0.577           &339.496        &106.484& 0.618           &321.624        &89.727         & 0.762           &198.214        &108.667        \\
      & -0.207          &-0.164         &0.446  & -0.420          &0.003          &0.314          & -0.243          &-0.244         &-0.776         \\
Cu2   & 0.459           &339.118        &104.214& 0.618           &343.403        &118.547        & 0.512           &180.205        &125.251        \\
      & -0.169          &-0.113         &0.356  & -0.170          &-0.295         &0.451          & -0.002          &-0.295         &-0.418         \\
Cu3   & 0.433           &161.513        &92.706 & 0.571           &185.893        &89.278         & 0.472           &236.501        &94.796         \\
      & 0.146           &-0.020         &-0.359 & -0.064          &0.007          &-0.594         & -0.423          &-0.040         &-0.417         \\
%% Editing table END
%%%%%% => R-factors for Na R:  6.51,  6.70,  5.32,  1.59
%%%%%% from fit file [/Users/pomjakushin/work/Fits/Albert/K2Cu3O(SO4)3/hrpt/fin/Na/Andrey/010/SA/afterSA/Na_1p7K-6K_mon1000_add50_P21c_C_varall_after_SA_fixed.sum]
%%%%%% => R-factors for NaK R:  6.57,  7.61,  7.03,  1.17
%%%%%% from fit file [/Users/pomjakushin/work/Fits/Albert/K2Cu3O(SO4)3/hrpt/fin/NaK/andrey/simann/new2023/afterSA/NaK_1p7K-6K_mon1000_add50_C2p_c_SYMM_basisOK_afterSA_fixed.sum]
\hline

\multicolumn{2}{l|}{$R_{p}, R_{wp}, R_{exp}, \chi^2$}&
\multicolumn{2}{l|}{6.45, 6.04, 4.65, 1.69}  & 
\multicolumn{3}{l|}{6.57,  7.61,  7.03,  1.17} &
\multicolumn{3}{l}{6.51,  6.70,    5.32,  1.59     } \\\hline % mY2-
%/Users/pomjakushin/work/Fits/Albert/K2Cu3O(SO4)3/hrpt/fin/K/Andrey_K/simann/new2023/afterSA/mk_table.pl
% Tue Oct 17 16:26:54 CEST 2023
% 
\multicolumn{10}{l}{ Least square fit (LSQ)}\\ \hline
%% Editing table BEGIN
Cu1   & 0.593(28)       &349(10)        &112.0(5.6)     & 0.617(53)       &325.6(9.9)     &93.1(6.0)      & 0.731(42)       &198.5(5.6)     &105.6(7.7)     \\
      & -0.1(1)         &-0.222(59)     &0.503(60)      & -0.4(1)         &-0.034(66)     &0.4(1)         & -0.241(75)      &-0.2(1)        &-0.757(43)     \\
Cu2   & 0.434(40)       &345(13)        &98.0(7.4)      & 0.581(83)       &350.2(9.3)     &119.4(6.2)     & 0.493(53)       &183.9(8.8)     &121.4(9.4)     \\
      & -0.1(1)         &-0.060(56)     &0.374(53)      & -0.095(89)      &-0.285(73)     &0.460(88)      & -0.031(68)      &-0.256(90)     &-0.431(32)     \\
Cu3   & 0.442(21)       &180            &90             & 0.532(51)       &188(11)        &87(26)         & 0.480(18)       &240.3(5.6)     &90             \\
      & 0               &0              &-0.442(21)     & -0.1(1)         &0.0(2)         &-0.561(70)     & -0.449(35)      &0              &-0.405(29)     \\
%% Editing table END
%%%%%% => R-factors for Na R:  6.57,  6.61,  5.19,  1.62
%%%%%% from fit file [/Users/pomjakushin/work/Fits/Albert/K2Cu3O(SO4)3/hrpt/fin/Na/Andrey/010/SA/afterSA/Na_1p7K-6K_mon1000_add50_P21c_C_varall_after_SA.sum]
%%%%%% => R-factors for NaK R:  6.64,  7.40,  6.84,  1.17
%%%%%% from fit file [/Users/pomjakushin/work/Fits/Albert/K2Cu3O(SO4)3/hrpt/fin/NaK/andrey/simann/new2023/afterSA/NaK_1p7K-6K_mon1000_add50_C2p_c_SYMM_basisOK_afterSA.sum]
%%%%%% => R-factors for K R:  6.47,  5.96,  4.55,  1.71
%%%%%% from fit file [/Users/pomjakushin/work/Fits/Albert/K2Cu3O(SO4)3/hrpt/fin/K/Andrey_K/simann/new2023/afterSA/K2_1p6K-6K_mon1000_add50_C2p_c_SYMM_after_SA.sum]
\hline

\multicolumn{2}{l|}{$R_{p}, R_{wp}, R_{exp}, \chi^2$}&
\multicolumn{2}{l|}{ 6.47,  5.96,  4.55,  1.71 }  &  % K
\multicolumn{3}{l|}{ 6.64,  7.40,  6.84,  1.17} &
\multicolumn{3}{l}{ 6.57,  6.61,  5.19,  1.62 } \\\hline

\multicolumn{10}{l}{ Least square fit (LSQ) Cu1 and Cu2 are constrained}\\ \hline
%%% Editing table BEGIN
Cu1   & 0.527(15)       &339.0(3.5)     &105.8(1.2)     & 0.574(25)       &339.7(7.8)     &104.8(3.0)     & 0.586(20)       &184.4(3.9)     &115.6(3.0)     \\
      & -0.194(31)      &-0.143(11)     &0.406(25)      & -0.210(77)      &-0.146(32)     &0.435(63)      & -0.043(39)      &-0.253(34)     &-0.543(20)     \\
Cu2   & 0.527(15)       &339.0(3.5)     &105.8(1.2)     & 0.574(25)       &339.7(7.8)     &104.8(3.0)     & 0.586(20)       &184.4(3.9)     &115.6(3.0)     \\
      & -0.194(31)      &-0.143(11)     &0.406(25)      & -0.210(77)      &-0.146(32)     &0.435(63)      & -0.043(39)      &-0.253(34)     &-0.543(20)     \\
Cu3   & 0.422(31)       &160.2(7.3)     &96(32)         & 0.526(53)       &192(11)        &97(53)         & 0.486(25)       &230.6(5.8)     &94(24)         \\
      & 0.152(52)       &-0.0(2)        &-0.342(44)     & -0.1(1)         &-0.1(5)        &-0.559(72)     & -0.404(41)      &-0.0(2)        &-0.458(26)     \\
%%% Editing table END
%%%%%% => R-factors for Na R:  6.84,  6.85,  5.22,  1.72
%%%%%% from fit file [/Users/pomjakushin/work/Fits/Albert/K2Cu3O(SO4)3/hrpt/fin/Na/Andrey/010/SA/afterSA/Na_1p7K-6K_mon1000_add50_P21c_C_varall_after_SA_Cu1=Cu2.sum]
%%%%%% => R-factors for NaK R:  6.76,  7.49,  6.90,  1.18
%%%%%% from fit file [/Users/pomjakushin/work/Fits/Albert/K2Cu3O(SO4)3/hrpt/fin/NaK/andrey/final_C2p_c/NaK_1p7K-6K_mon1000_add50_C2p_c_SYMM_varall_after_SA_a_basisOK_Cu1=Cu2.sum]
%%%%%% => R-factors for K R:  6.47,  5.93,  4.57,  1.69
%%%%%% from fit file [/Users/pomjakushin/work/Fits/Albert/K2Cu3O(SO4)3/hrpt/fin/K/Andrey_K/simann/new2023/afterSA/K2_1p6K-6K_mon1000_add50_C2p_c_SYMM_after_SA_Cu1=Cu2.sum]
\hline

\multicolumn{2}{l|}{$R_{p}, R_{wp}, R_{exp}, \chi^2$}&
\multicolumn{2}{l|}{ 6.76,  7.49,  6.90,  1.18 }  &  % K
\multicolumn{3}{l|}{ 6.76,  7.49,  6.90,  1.18 }  &  % NaK
 \multicolumn{3}{l}{  6.84,  6.85,  5.22,  1.72 } \\\hline % Na

%%%% This is for table
%\end{tabular}
%\end{table}
%%%% --------------------
%\end{center}
\end{longtable}
\clearpage  % just to stop any pending floats landing on the same page

%\begin{table}
%
%\caption{Magnetic model parameters for \acuos\ (Na) refined from
% The numeration of the
%  atoms is the same as in Table~\ref{magtab}. $M$ is the size of the
%  magnetic moment, $\phi$ and $\theta$ are spherical angles with $c$
%  and $b$ axes, respectively. The errorbars are given only for the
%  independently refined parameters. 
%}
%
%\label{magmom2}
%
%\begin{center}
%\begin{tabular}{l|l|l|l}
%
%\multicolumn{3}{c|}{K} \\\hline
% &$M, \mu_B$, $m_x$       & $\phi$, $m_y$     & $\theta$, $m_z$ 
% \\ \hline
%\multicolumn{10}{l}{ Simulated annealing (SA)}\\ \hline
%Cu1  & 0.577	&339.496	&106.484 \\
%     & 0.207	&0.164		&0.446   \\
%Cu2  & 0.459	&339.118	&104.214 \\
%     & 0.169	&0.113		&0.356   \\
%Cu3  & 0.433	&161.513	&92.706  \\
%     & 0.146	&0.020		&0.359   \\ \hline
%
%\end{tabular}
%\end{center}
%\end{table}

\def\extgra{pdf}
\def\figsiz{\textwidth}
\def\figsiz{8cm}
\def\figsizcm{8}

%===========================      ========================
%                           Figures
%===========================      ========================

\begin{figure}
  \begin{center}
    \includegraphics[width=\figsiz]{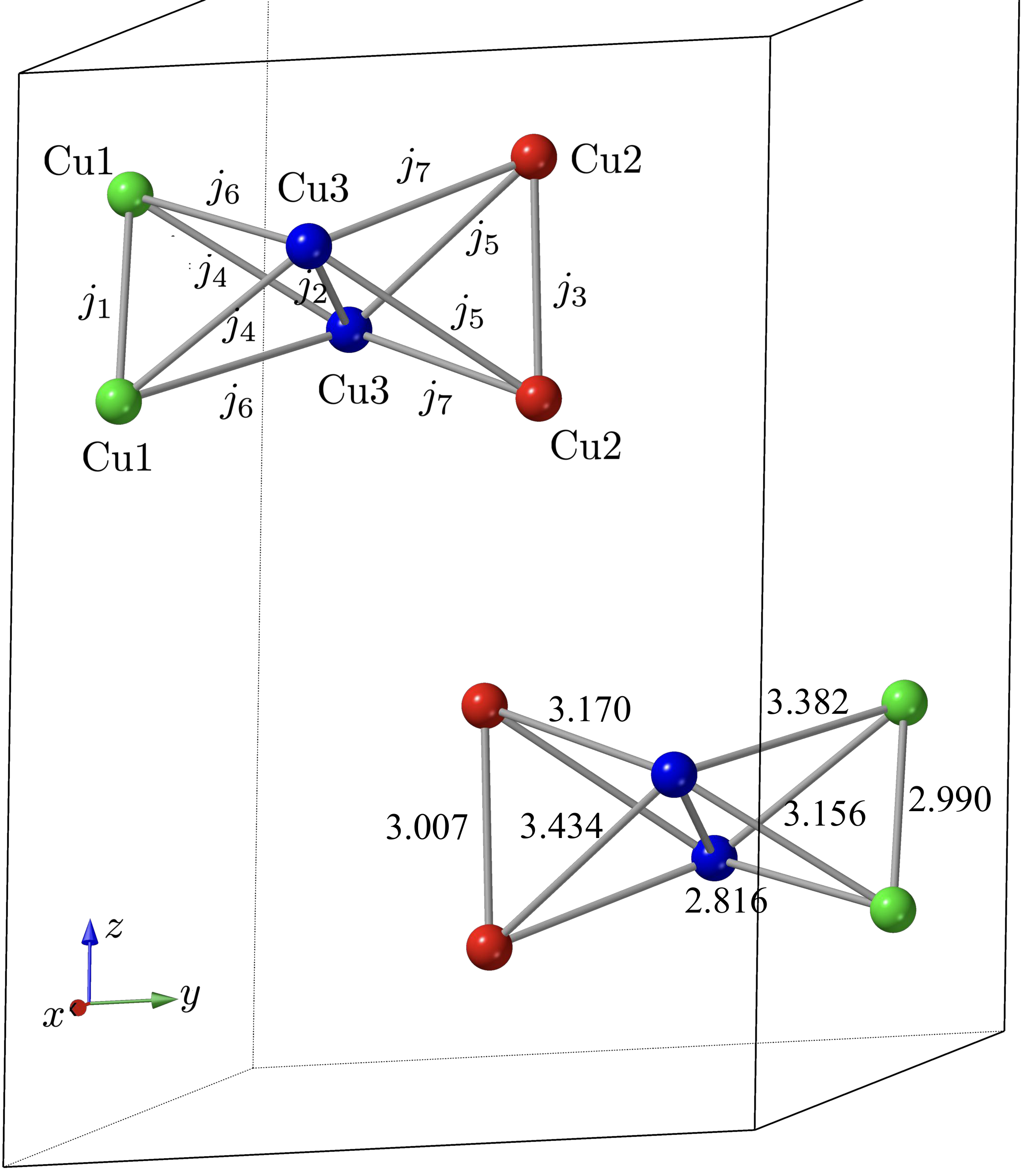} %{dpcu2ni.\extgra}
  \end{center}

  \caption{Two hexamers in NaK-sample out of four in the unit cell. The hexamers are related by inversion. Two other hexamers, which are not shown are related by C-centering translation (1/2,1/2,0). The bond lengths in [\AA] are indicated for the bottom hexamer. $J_i$ ($i=1..7$) show the exchange coupling constants used in section~\ref{triplet} and in Formula~\ref{Hamiltonian}.  }
  \label{hexamer}
\end{figure}

\begin{figure}
  \begin{center}
    \includegraphics[width=\figsiz]{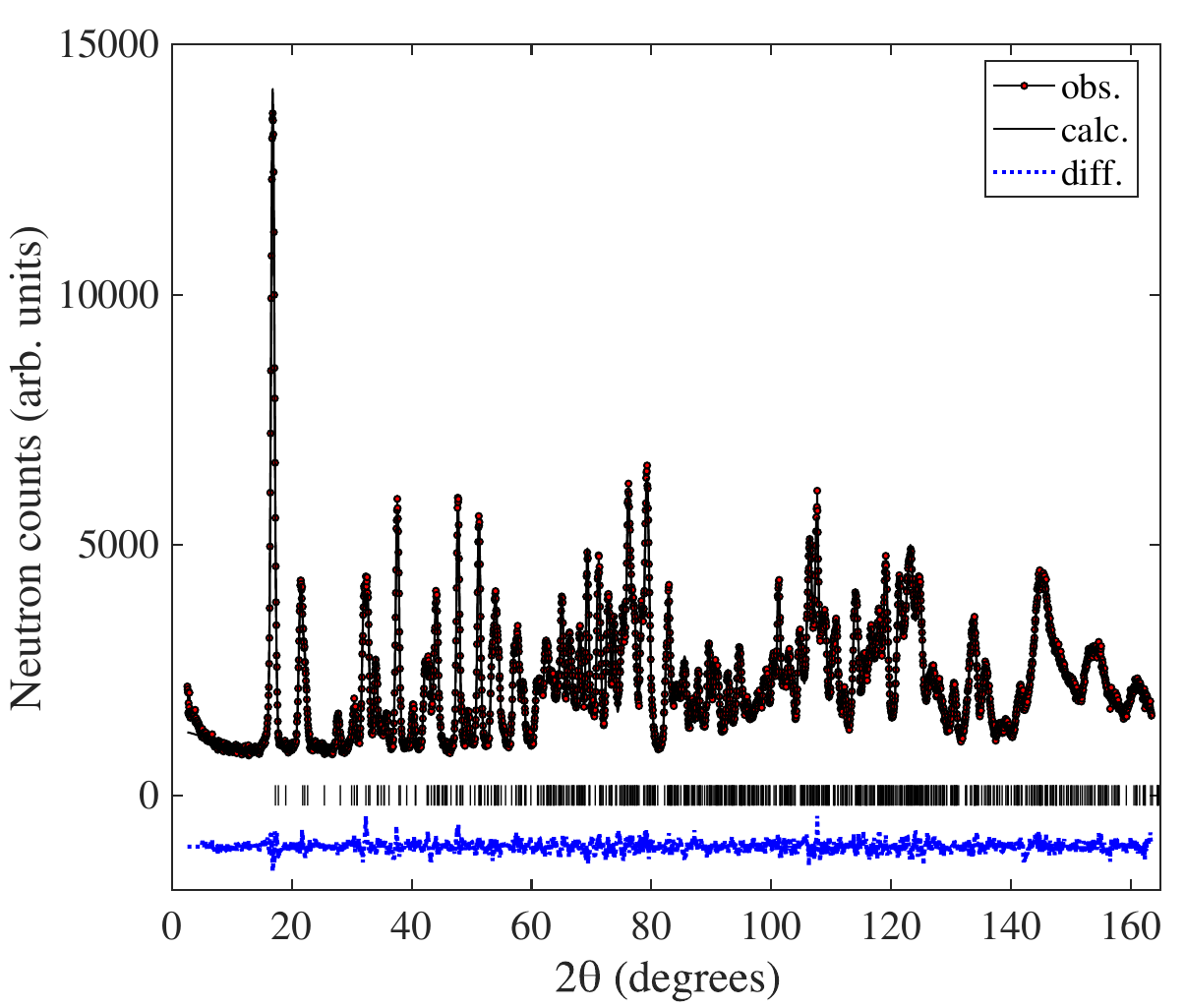} %{dpcu2ni.\extgra}
  \end{center}

  \caption{The Rietveld refinement pattern and difference plot of the
    neutron diffraction data for the sample \acuos\ (A$_2$=NaK) at T=1.6~K
    measured at HRPT with the wavelength $\lambda=2.45$~\AA. The rows
    of tics show the Bragg peak positions. The difference between observed and calculated intensities is shown by the dotted blue line. }
 
  \label{dif_HRPT}
\end{figure}

\begin{figure}
  \begin{center}
    \includegraphics[width=\figsiz]{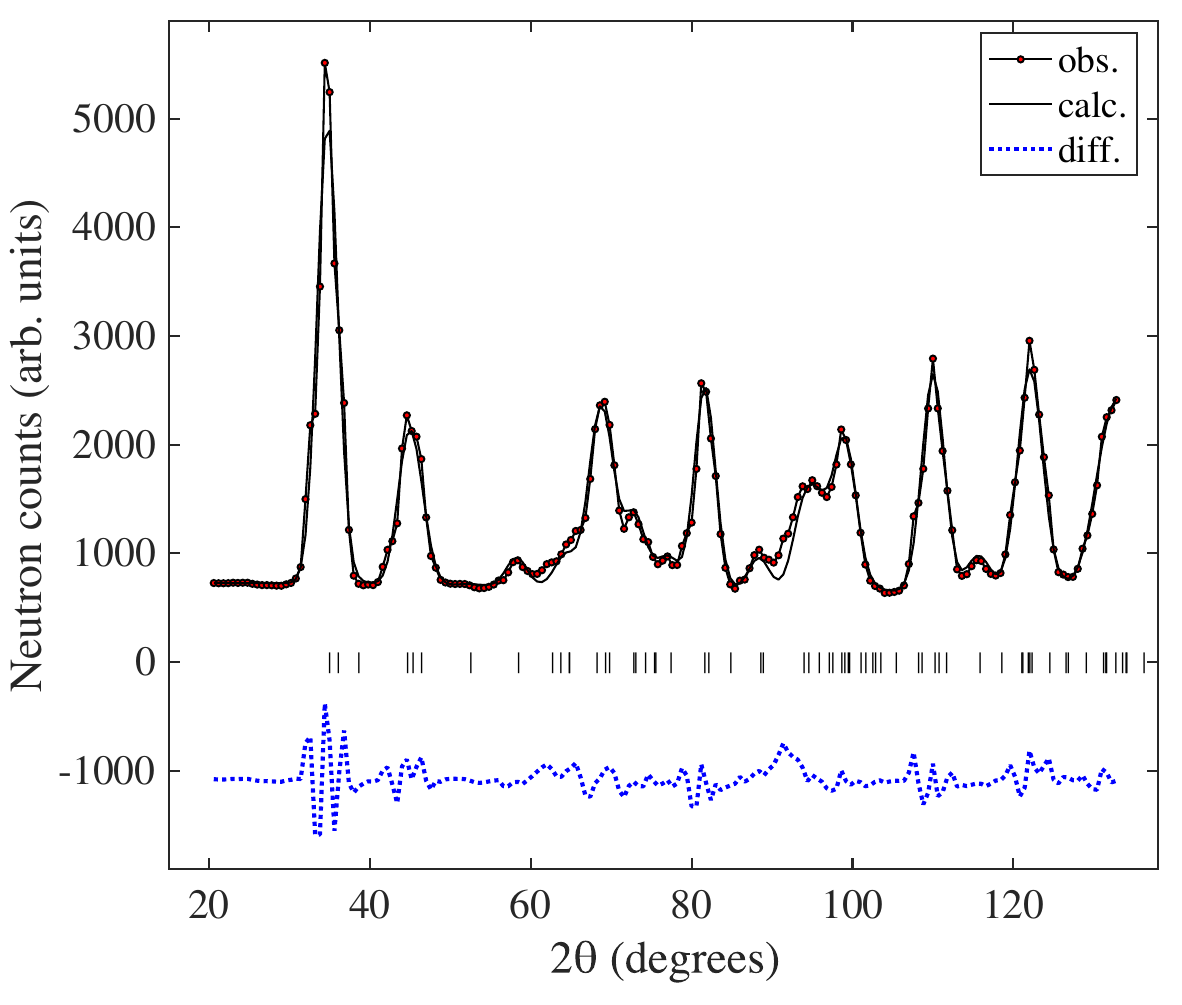} %{dpcu2ni.\extgra}
  \end{center}

 \caption{The Rietveld refinement pattern and difference plot of the
    neutron diffraction data for the sample \acuos\ (A$_2$=NaK) at T=6~K
    CNCS/SNS with the wavelength $\lambda=4.96$~\AA. All structure parameters were fixed by the values obtained from HRPT data (Fig.~\ref{dif_HRPT}). The rows
    of tics show the Bragg peak positions. The difference between observed and calculated intensities is shown by the dotted blue line. See the text for details.}

  \label{dif_CNCS}
\end{figure}

\begin{figure}
  \begin{center}
    \includegraphics[width=\figsiz]{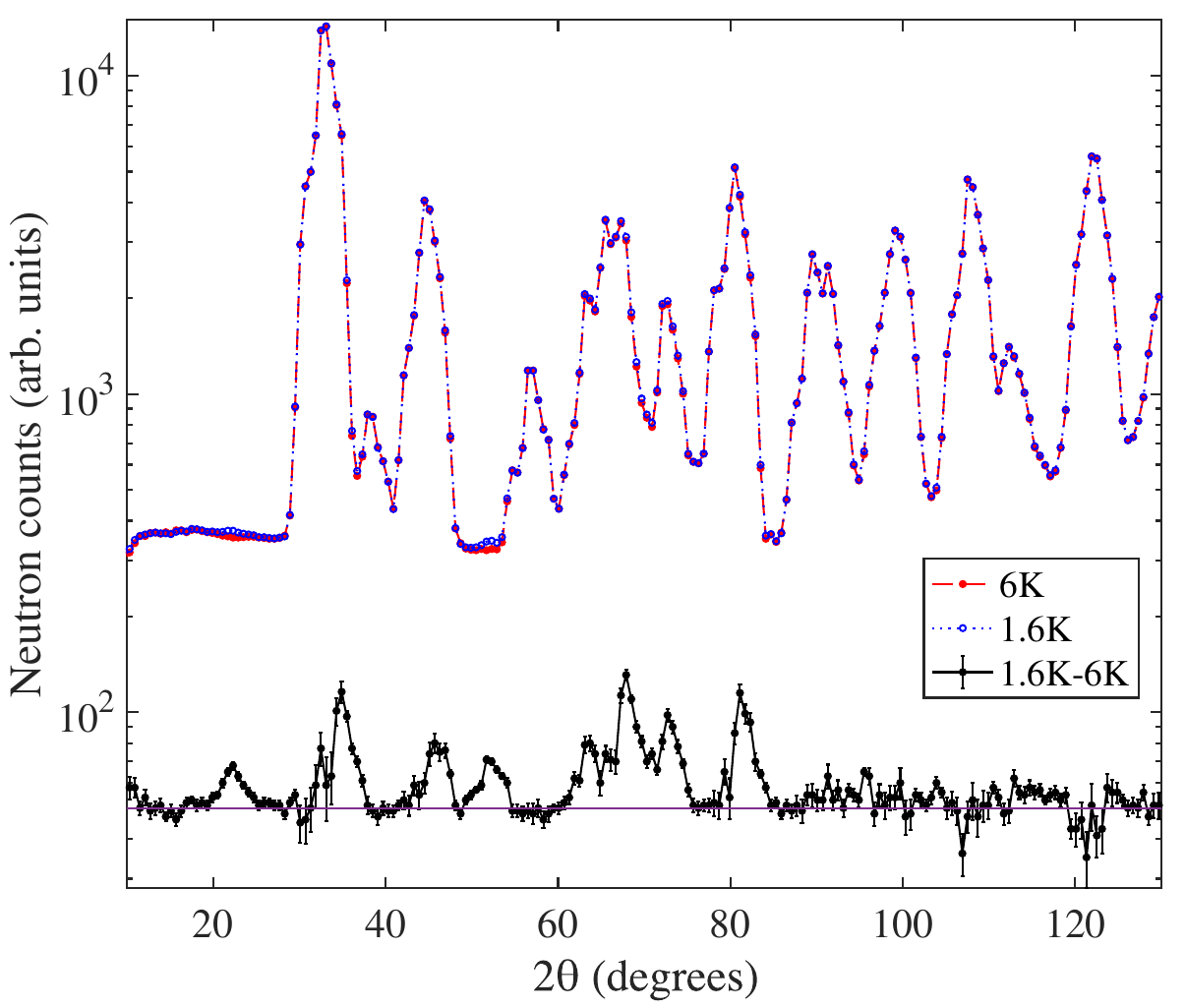} %{dpcu2ni.\extgra}
  \end{center}

  \caption{The raw neutron diffraction patterns for the sample \acuos\ (A=K) at T=1.6~K (blue dotted line and open symbols), 6~K (red dashed line and closed symbols) and difference pattern containing purely magnetic contribution (black line and closed symbols), measured at CNCS/SNS with the wavelength $\lambda=4.96$~\AA.  A constant 50 has been added to the difference pattern. Note, the $y$-axis has logarithmic scale. }
  \label{dif_raw}
\end{figure}

\begin{figure}
  \begin{center}
    \includegraphics[width=\figsiz]{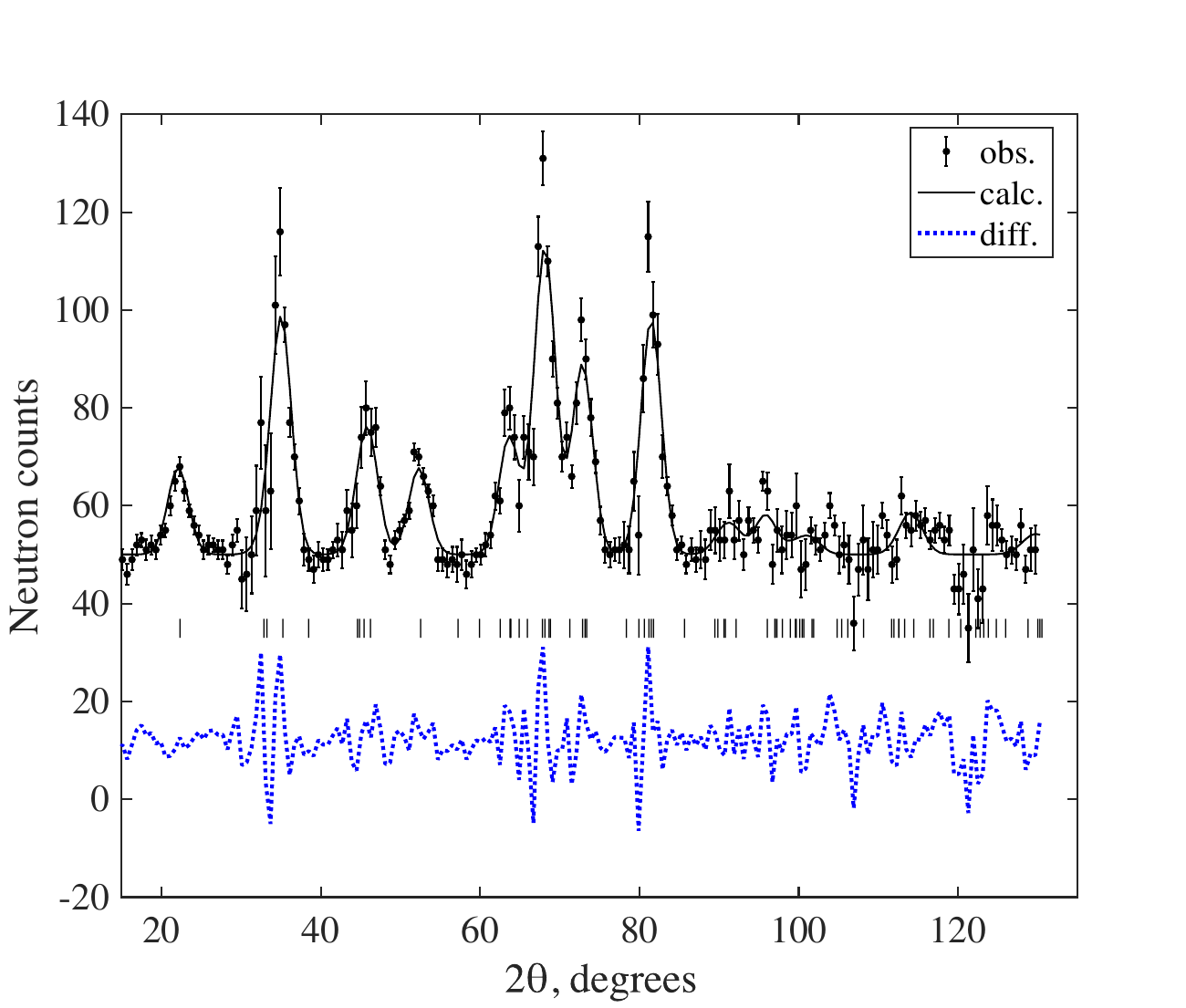} %{dpcu2ni.\extgra}
  \end{center}

  \caption{Model free le Bail fit of the difference pattern for the sample \acuos\ (A=K) containing purely magnetic contribution, measured at CNCS/SNS with the wavelength $\lambda=4.96$~\AA. The rows of tics show the Bragg peak positions. The difference between observed and calculated intensities is shown by the dotted blue line. The reliability factors~\cite{Fullprof} $R_{p}$, $R_{wp}$, $R_{exp}$, $\chi^2$ are 6.15, 5.88, 4.65, 1.60. }
  \label{dif_leBail_K}
\end{figure}

\begin{figure}
  \begin{center}
    \includegraphics[width=\figsiz]{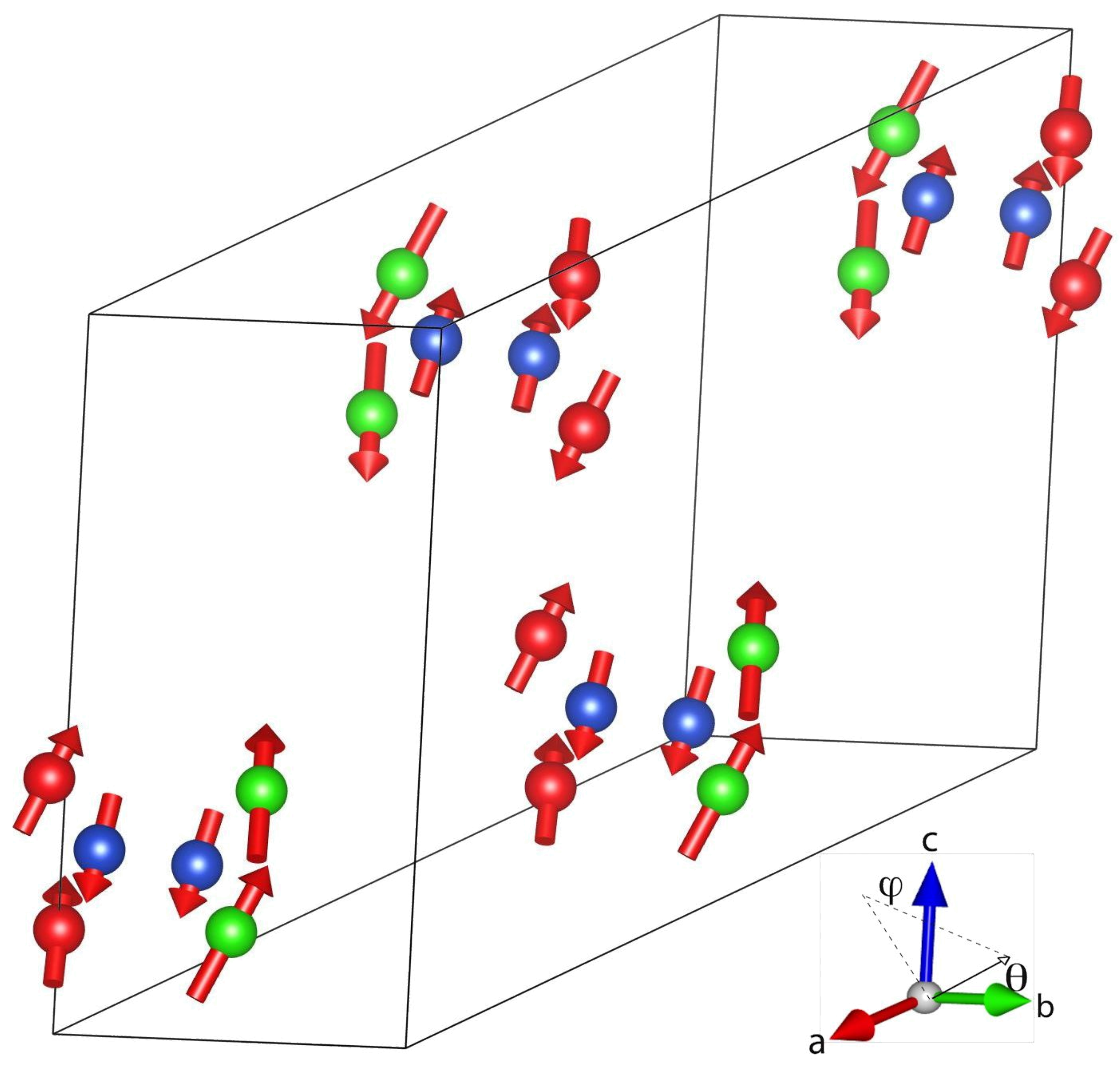}
    \includegraphics[width=\figsiz]{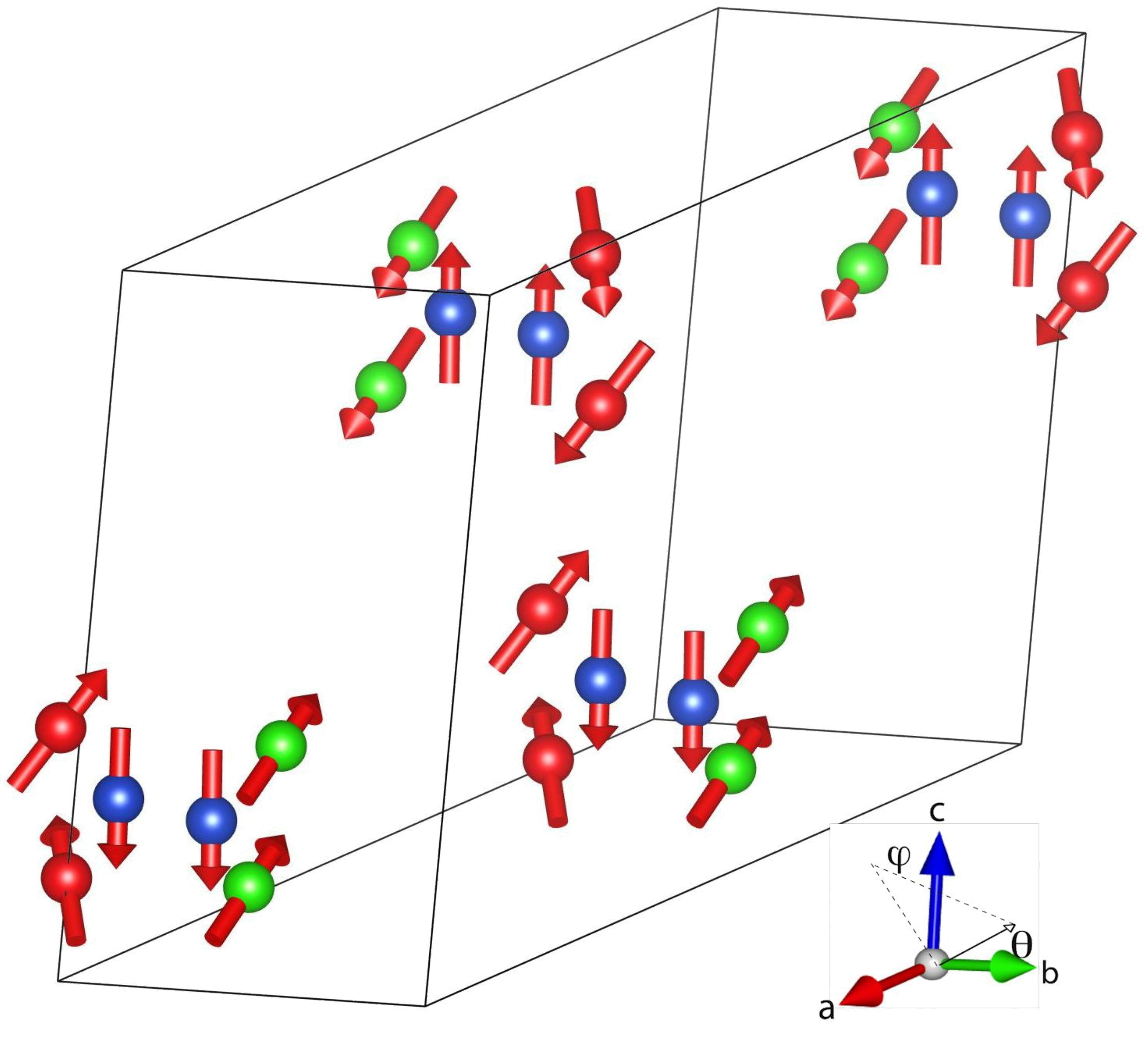}
    \includegraphics[width=\figsiz]{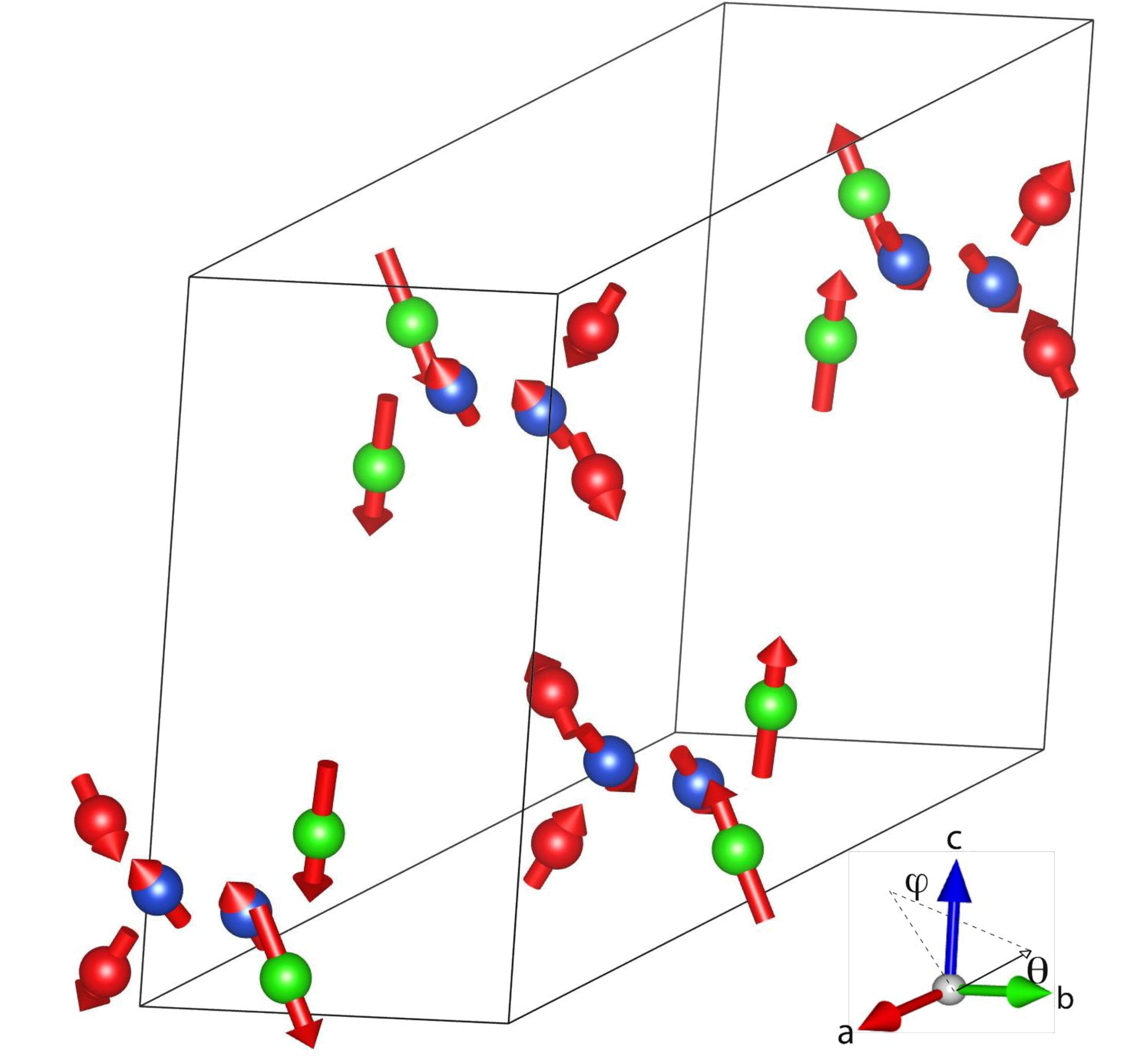}    
  \end{center}

  \caption{Magnetic structures formed by hexamers in magnetic space groups MSG (top) C2'/c and (bottom) P2\_1/c.1'\_C[C2/c] (UNI symbol~\cite{Campbell_UNI}), or C\_P2'/c (OG-symbol). The number of atoms is 24 in one unit cell, but some atoms are shown outside the unit cell to better show the four hexamers. The hexamers related by inversion in C2'/c settings along ($bc$)-diagonal (those that are closest and that are arranged approximately in vertical direction in the figure) are coupled AFM. Two other hexamers, are related by C-centering translation (1/2,1/2,0) are coupled FM for the samples with A$_2$ =K$_2$ and NaK  (left and right top structures), but AFM for the sample with A=Na (bottom). Side spin pairs Cu1 and Cu2 are in green and red, the central Cu3 pair is in blue.}
  \label{mag_strs}
\end{figure}

%/Users/pomjakushin/work/Fits/Albert/K2Cu3O(SO4)3/hrpt/fin/NaK/andrey/final_C2p_c/NaK_1p7K-6K_mon1000_add50_C2p_c_SYMM_varall_after_SA_a_basisOK.prf
% (100) 22.43
\begin{figure}
  \begin{center}
    \includegraphics[width=\figsiz]{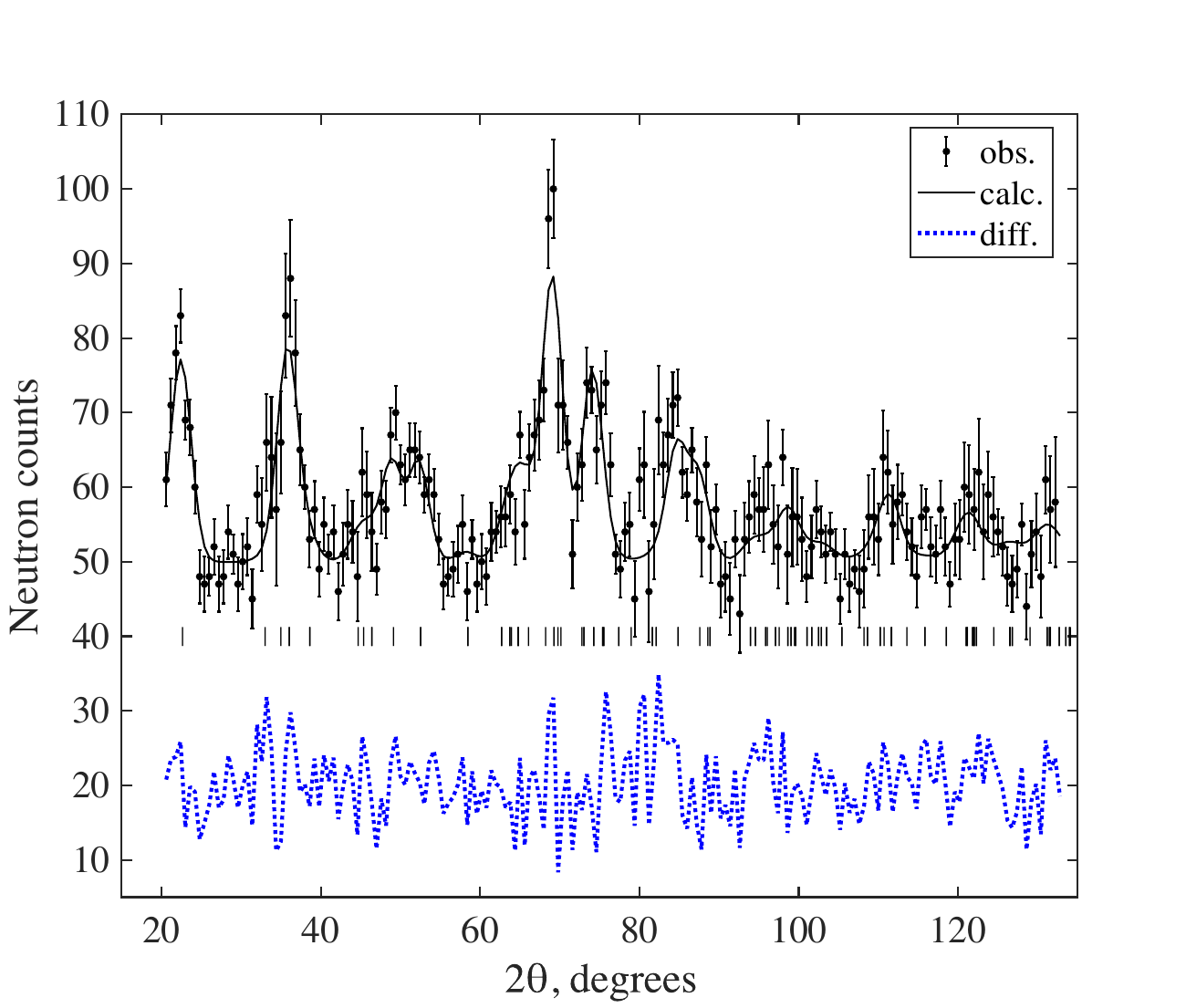} %{dpcu2ni.\extgra}
  \end{center}

  \caption{Fit to C2'/c-model of the difference pattern for the sample \acuos\ (A$_2$=NaK) containing purely magnetic contribution, measured at CNCS/SNS with the wavelength $\lambda=4.96$~\AA. The rows of tics show the Bragg peak positions. The difference between observed and calculated intensities is shown by the dotted blue line.}
  \label{dif_mag_NaK}
\end{figure}

%'/Users/pomjakushin/work/Fits/Albert/K2Cu3O(SO4)3/hrpt/fin/K/Andrey_K/simann/K2_1p6K-6K_mon1000_add50_C2p_c_SYMM_varall_after_SA.prf'...
% 22.09
\begin{figure}
  \begin{center}
    \includegraphics[width=\figsiz]{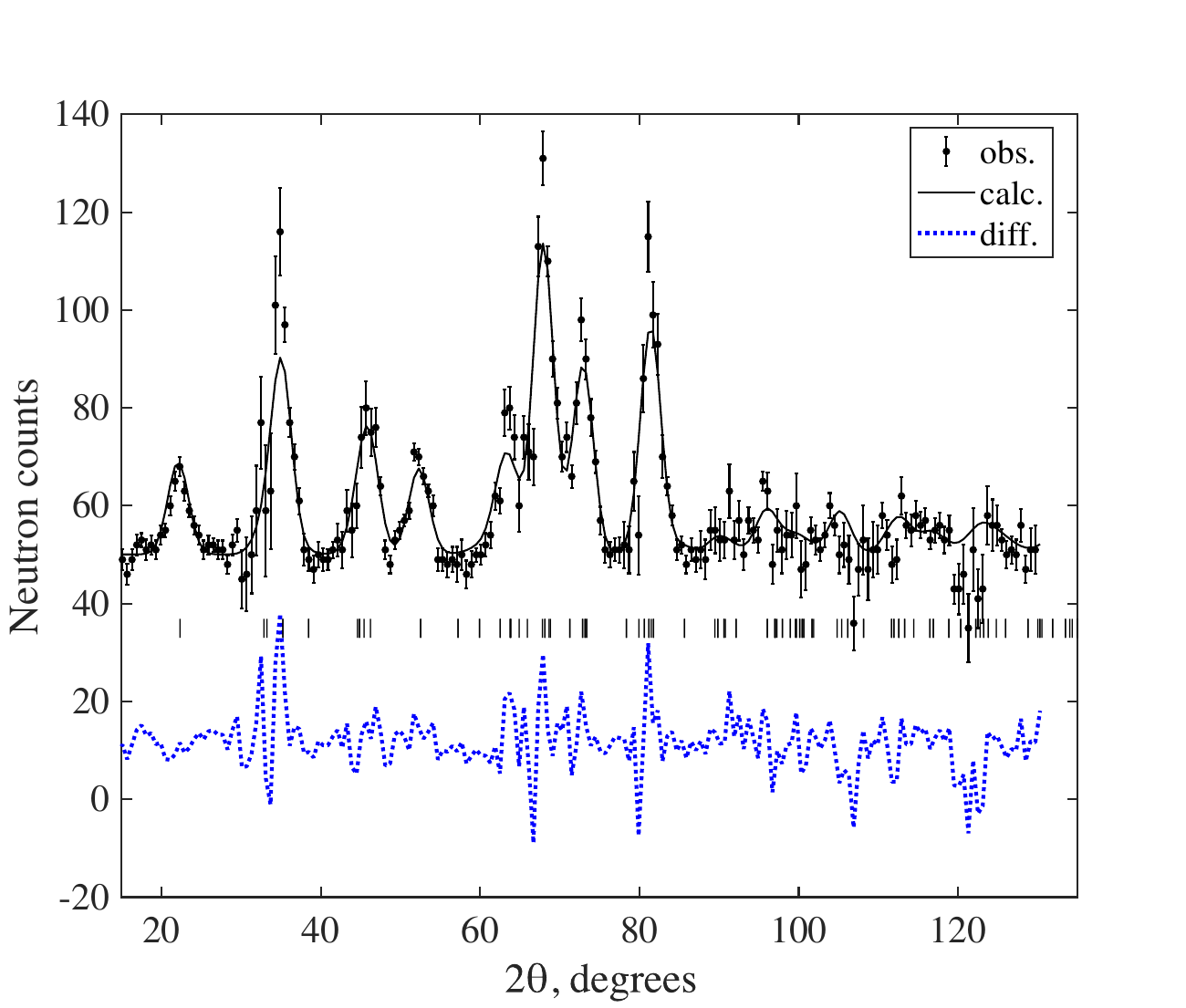} %{dpcu2ni.\extgra}
  \end{center}

  \caption{Fit to C2'/c-model of the difference pattern for the sample \acuos\ (A=K) containing purely magnetic contribution, measured at CNCS/SNS with the wavelength $\lambda=4.96$~\AA. The rows of tics show the Bragg peak positions. The difference between observed and calculated intensities is shown by the dotted blue line. The reliability factors~\cite{Fullprof} $R_{p}$, $R_{wp}$, $R_{exp}$, $\chi^2$ are 6.45, 6.04, 4.65, 1.69.}
  \label{dif_mag_K}
\end{figure}

%%%/Users/pomjakushin/work/Fits/Albert/K2Cu3O(SO4)3/hrpt/fin/Na/Andrey/simann/rietveld/Na_1p7K-6K_mon1000_add50_C2p_c_SYMM_varall_after_SA_a_basisOK.prf
%%% (100) 21.8
%%\begin{figure}
%%  \begin{center}
%%  \end{center}
%%
%%  \caption{Fit to C2'/c-model of the difference pattern for the sample \acuos\ (A=Na) containing purely magnetic contribution, measured at CNCS/SNS with the wavelength $\lambda=4.96$~\AA.}
%%  \label{dif_mag_Na}
%%\end{figure}
%/Users/pomjakushin/work/texts/Papers/K2Cu3O(SO4)3/figures_maybe/matlab/plot_prf_mag_Na_mY2m.m
\begin{figure}
  \begin{center}
    \includegraphics[width=\figsiz]{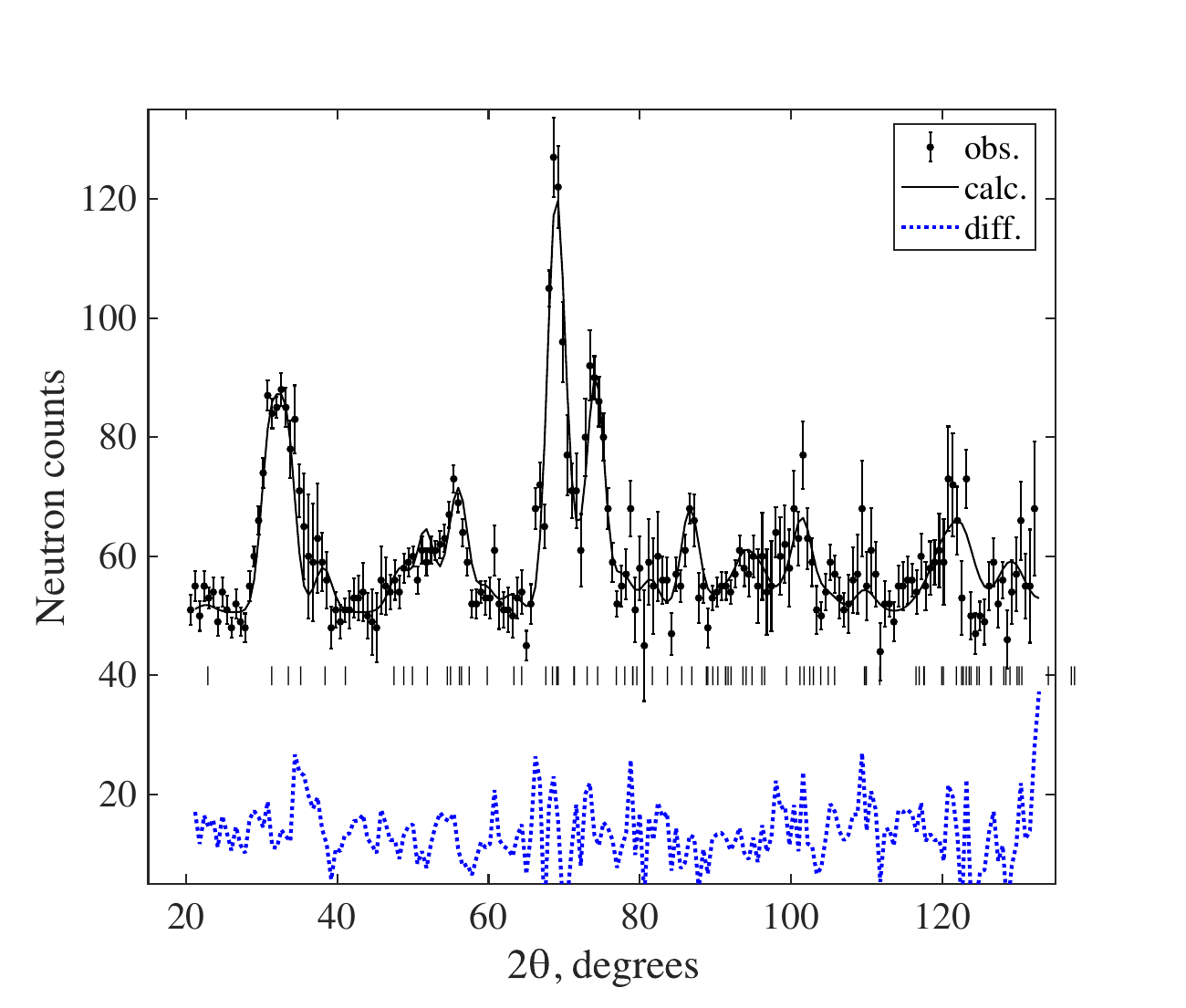} %{dpcu2ni.\extgra}
  \end{center}

  \caption{Fit to the  model P2\_1/c.1'\_C[C2/c] (basis transformation from parent %$\vec{A}=\vec{a}$, $\vec{B}=\vec{b}$, $\vec{C}=\vec{c}$; 1/4,1/4,0) of the difference pattern
(1,0,0), (0,1,0), (0,0,1); 1/4,1/4,0) of the difference pattern for the sample \acuos\ (A=Na) containing purely magnetic contribution, measured at CNCS/SNS with the wavelength $\lambda=4.96$~\AA. The rows of tics show the Bragg peak positions. The difference between observed and calculated intensities is shown by the dotted blue line.}
  \label{dif_mag_Na}
\end{figure}

% simple math, arithmetics \advance
% https://en.wikibooks.org/wiki/TeX#TeX_Primitives
% https://mirror.init7.net/ctan/info/impatient/book.pdf
% \count<number>=<integer>
% \countdef\<name>=<0..255>
\count1=\figsizcm
%\count1=8
\multiply\count1 by 4
\divide\count1 by 5
\begin{figure}
  \begin{center}
    \includegraphics[width=\figsiz]{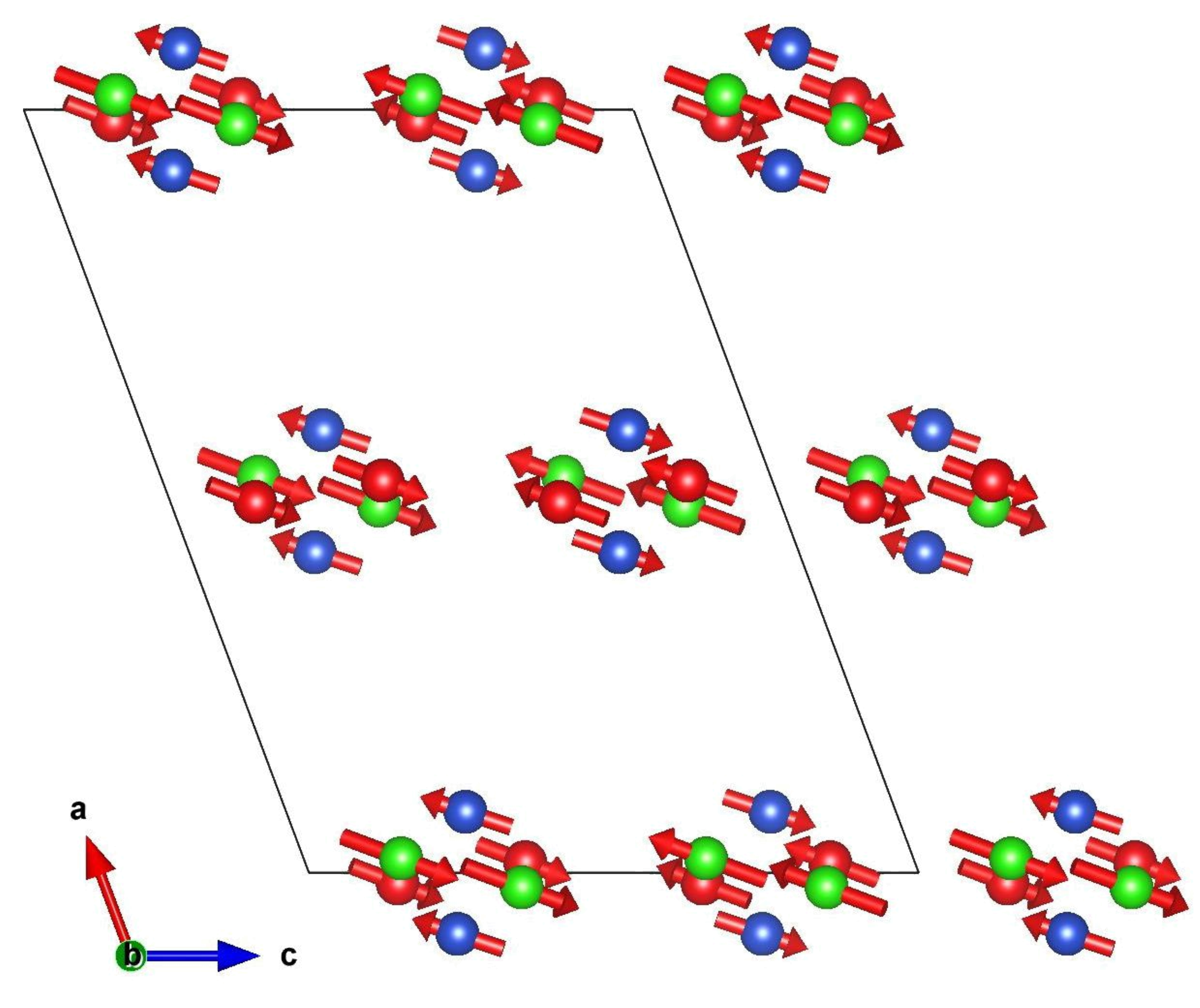}
    \includegraphics[width=\figsiz]{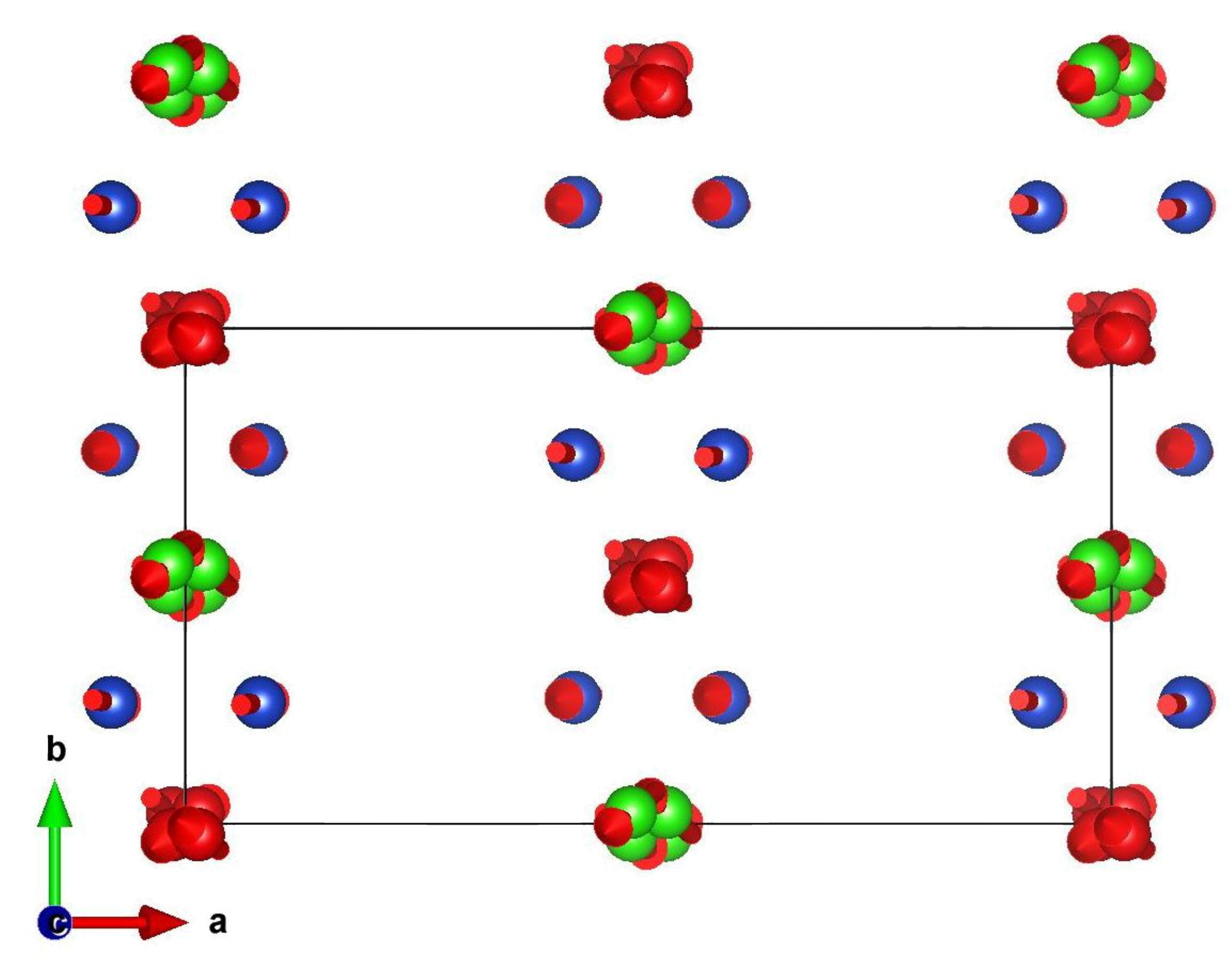}
    \includegraphics[width=\count1cm]{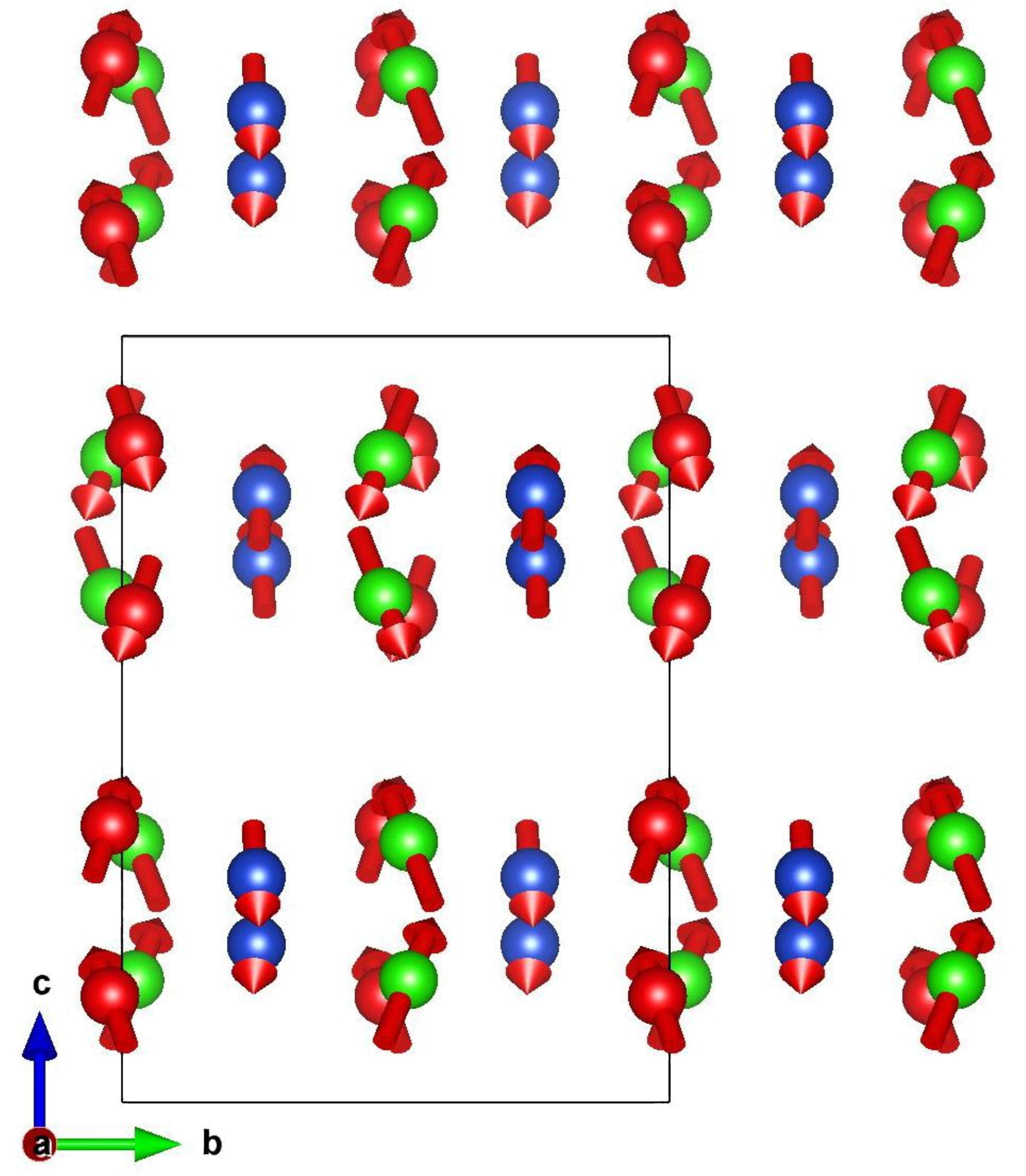}    
  \end{center}

  \caption{Magnetic structure in the sample with A=K (C2'/c) in three projections. $ac$ and $bc$ layers are FM stacked along $b$- and $a$-axis directions, respectively. Side spin pairs Cu1 and Cu2 are in green and red, the central Cu3 pair is in blue. }
  \label{K_mag_projections}
\end{figure}

\count1=\figsizcm
\multiply\count1 by 3
\divide\count1 by 2

%\end{document}

\clearpage
{
\section{Supplementary Material}

Three magnetic crystallographic information files (mcif) are given below.

\renewcommand{\baselinestretch}{0.75}
\begin{verbatim}

#############################################
#
# File: Khexamers_CM.mcif
#
#############################################

_parent_space_group.name_H-M_alt  'C 2/c'
_parent_space_group.IT_number      15
_parent_space_group.transform_Pp_abc  'a,b,c;0,0,0'
_parent_space_group.child_transform_Pp_abc  'a,b,c;0,0,0'

loop_
_parent_propagation_vector.id
_parent_propagation_vector.kxkykz
k1 [0 0 0]
# mGM2-

_cell_length_a                  18.975500
_cell_length_b                   9.500400
_cell_length_c                  14.197200
_cell_angle_alpha               90.000000
_cell_angle_beta               110.491500
_cell_angle_gamma               90.000000

_space_group_magn.number_BNS "15.87"
_space_group_magn.name_UNI "C2'/c"
_space_group_magn.name_BNS "C2'/c"
_space_group_magn.number_OG "15.3.94"
_space_group_magn.name_OG "C2'/c"
_space_group_magn.point_group_number "5.3.14"
_space_group_magn.point_group_name_UNI "2'/m"

loop_
_space_group_magn_transforms.id
_space_group_magn_transforms.Pp_abc
_space_group_magn_transforms.source
1 a,b,c;0,0,0 "BNS"
2 a,b,c;0,0,0 "OG"

loop_
_space_group_symop_magn_operation.id
_space_group_symop_magn_operation.xyz
1 x,y,z,+1 
2 x,-y,z+1/2,+1 
3 -x,y,-z+1/2,-1 
4 -x,-y,-z,-1 

loop_
_space_group_symop_magn_centering.id
_space_group_symop_magn_centering.xyz
1 x,y,z,+1 
2 x+1/2,y+1/2,z,+1 

loop_
_atom_site_label
_atom_site_type_symbol
_atom_site_occupancy
_atom_site_fract_x
_atom_site_fract_y
_atom_site_fract_z
_atom_site_U_iso_or_equiv
 Cu1  Cu 1.0  0.4813  0.0226  0.3403  0.00140
 Cu2  Cu 1.0  0.4858  0.4778  0.1392  0.00140
 Cu3  Cu 1.0  0.4199  0.7451  0.2055  0.00140
 K1  K 1.0  0.3283  0.7559  0.4415  .  
 K2  K 1.0  0.1981  0.7433  0.1272  .  
 O1  O 1.0  0.5  0.8876  0.25  0.01030
 O2  O 1.0  0.4506  0.8262  0.4062  0.01030
 O3  O 1.0  0.5631  0.6792  0.4564  0.01030
 O4  O 1.0  0.466  0.6446  0.531  0.01030
 O5  O 1.0  0.5899  0.0557  0.4109  0.01030
 O6  O 1.0  0.4061  0.4422  0.3233  0.01030
 O7  O 1.0  0.3378  0.6189  0.2031  0.01030
 O8  O 1.0  0.2806  0.3953  0.1982  0.01030
 O9  O 1.0  0.5488  0.8516  0.5703  0.01030
 O10  O 1.0  0.6201  0.0687  0.2548  0.01030
 O11  O 1.0  0.3853  0.4109  0.145  0.01030
 O12  O 1.0  0.5  0.6051  0.25  0.01030
 O13  O 1.0  0.6587  0.8724  0.3682  0.01030
 O14  O 1.0  0.7166  0.0973  0.4246  0.01030
 S1  S 1.0  0.5103  0.7576  0.4879  .  
 S2  S 1.0  0.6479  0.0163  0.3672  .  
 S3  S 1.0  0.3535  0.4633  0.2158  .  

loop_
_atom_site_moment.label
_atom_site_moment.crystalaxis_x
_atom_site_moment.crystalaxis_y
_atom_site_moment.crystalaxis_z
_atom_site_moment.symmform
Cu1 -0.1(1) -0.222(59)  0.504(61)  Mx,My,Mz
Cu2 -0.1(1) -0.060(56)  0.374(54)  Mx,My,Mz
Cu3 0 0  -0.442(21)  Mx,My,Mz

# end of mcif

#############################################
#
# File: NaKhexamers_CM.mcif
#
#############################################

_parent_space_group.name_H-M_alt  'C 2/c'
_parent_space_group.IT_number      15
_parent_space_group.transform_Pp_abc  'a,b,c;0,0,0'
_parent_space_group.child_transform_Pp_abc  'a,b,c;0,0,0'

loop_
_parent_propagation_vector.id
_parent_propagation_vector.kxkykz
k1 [0 0 0]
# mGM2-


_cell_length_a                  18.473000
_cell_length_b                   9.364400
_cell_length_c                  14.314600
_cell_angle_alpha               90.000000
_cell_angle_beta               113.964200
_cell_angle_gamma               90.000000

_space_group_magn.number_BNS "15.87"
_space_group_magn.name_UNI "C2'/c"
_space_group_magn.name_BNS "C2'/c"
_space_group_magn.number_OG "15.3.94"
_space_group_magn.name_OG "C2'/c"
_space_group_magn.point_group_number "5.3.14"
_space_group_magn.point_group_name_UNI "2'/m"

loop_
_space_group_magn_transforms.id
_space_group_magn_transforms.Pp_abc
_space_group_magn_transforms.source
1 a,b,c;0,0,0 "BNS"
2 a,b,c;0,0,0 "OG"

loop_
_space_group_symop_magn_operation.id
_space_group_symop_magn_operation.xyz
1 x,y,z,+1 
2 x,-y,z+1/2,+1 
3 -x,y,-z+1/2,-1 
4 -x,-y,-z,-1 

loop_
_space_group_symop_magn_centering.id
_space_group_symop_magn_centering.xyz
1 x,y,z,+1 
2 x+1/2,y+1/2,z,+1 


loop_
_atom_site_label
_atom_site_type_symbol
_atom_site_occupancy
_atom_site_fract_x
_atom_site_fract_y
_atom_site_fract_z
_atom_site_U_iso_or_equiv
 Cu1  Cu 1.0  0.4827  0.0205  0.343  0.00830
 Cu2  Cu 1.0  0.4834  0.4724  0.1381  0.00830
 Cu3  Cu 1.0  0.4166  0.7483  0.2043  0.00830
 K1  K 1.0  0.3245  0.7595  0.4573  0.01300
 Na1  Na 1.0  0.1938  0.7399  0.1272  0.01300
 O1  O 1.0  0.5  0.8907  0.25  0.00840
 O2  O 1.0  0.4382  0.8272  0.3963  0.00840
 O3  O 1.0  0.5577  0.6727  0.4592  0.00840
 O4  O 1.0  0.4546  0.6515  0.5186  0.00840
 O5  O 1.0  0.5961  0.0596  0.4144  0.00840
 O6  O 1.0  0.4011  0.436  0.3144  0.00840
 O7  O 1.0  0.3314  0.6087  0.1864  0.00840
 O8  O 1.0  0.2718  0.3869  0.1872  0.00840
 O9  O 1.0  0.5413  0.8558  0.5656  0.00840
 O10  O 1.0  0.6228  0.0826  0.2608  0.00840
 O11  O 1.0  0.3807  0.3986  0.1372  0.00840
 O12  O 1.0  0.5  0.5992  0.25  0.00840
 O13  O 1.0  0.6698  0.8755  0.3727  0.00840
 O14  O 1.0  0.7285  0.1005  0.4306  0.00840
 S1  S 1.0  0.4957  0.7514  0.4873  0.00800
 S2  S 1.0  0.6567  0.0345  0.3771  0.00800
 S3  S 1.0  0.3486  0.4431  0.2126  0.00800

Cu1 -0.4(1) -0.034(66)  0.4(1)  Mx,My,Mz
Cu2 -0.095(89) -0.285(73)  0.460(88)  Mx,My,Mz
Cu3 -0.1(1) 0.0(2)  -0.561(70)  Mx,My,Mz

# end of mcif

#############################################
#
# File: Nahexamers_ISO.mcif
#
#############################################

# Space Group:  15 C2/c       C2h-6
# Default space-group preferences: monoclinic axes a(b)c, monoclinic cell choice 1, orthorhombic axes abc, origin choice 2, hexagonal axes, SSG standard setting
# Lattice parameters: a=17.21410, b=9.37287, c=14.37010, alpha=90.00000, beta=111.84400, gamma=90.00000
# k point: Y, k8 (0,1,0)
# IR: mY2-, mk8t4
#  P1   (a) 14.84 P2_1/c.1'_C[C2/c], basis={(1,0,0),(0,1,0),(0,0,1)}, origin=(1/4,1/4,0), s=2, i=2, k-active= (0,1,0)

_parent_space_group.name_H-M_alt  'C 2/c'
_parent_space_group.IT_number      15
_parent_space_group.transform_Pp_abc  'a,b,c;0,0,0'
_parent_space_group.child_transform_Pp_abc  'a,b,c;1/4,1/4,0'

loop_
_parent_propagation_vector.id
_parent_propagation_vector.kxkykz
k1 [0 1 0]
# IR: mY2-, mk8t4

_cell_length_a    17.21410
_cell_length_b    9.37287
_cell_length_c    14.37010
_cell_angle_alpha 90.00000
_cell_angle_beta  111.84400
_cell_angle_gamma 90.00000
_cell_volume      2152.08014

_space_group_magn.number_BNS "14.84"
_space_group_magn.name_UNI "P2_1/c.1'_C[C2/c]"
_space_group_magn.name_BNS "P_C2_1/c"
_space_group_magn.number_OG "15.7.98"
_space_group_magn.name_OG "C_P2'/c"
_space_group_magn.point_group_number "5.2.13"
_space_group_magn.point_group_name_UNI "2/m.1'"

loop_
_space_group_magn_transforms.id
_space_group_magn_transforms.Pp_abc
_space_group_magn_transforms.source
1 a,b,c;0,0,0 "BNS"
2 a,b,c;0,0,1/2 "OG"

loop_
_space_group_symop_magn_operation.id
_space_group_symop_magn_operation.xyz
1 x,y,z,+1 
2 -x,y+1/2,-z+1/2,+1 
3 -x,-y,-z,+1 
4 x,-y+1/2,z+1/2,+1 

loop_
_space_group_symop_magn_centering.id
_space_group_symop_magn_centering.xyz
1 x,y,z,+1 
2 x+1/2,y+1/2,z,-1 

loop_
_atom_site_label
_atom_site_type_symbol
_atom_site_symmetry_multiplicity
_atom_site_Wyckoff_label
_atom_site_fract_x
_atom_site_fract_y
_atom_site_fract_z
_atom_site_occupancy
_atom_site_fract_symmform
Cu1 Cu   8 f 0.22740  0.77021  0.34063 1.00000 Dx,Dy,Dz 
Cu2 Cu   8 f 0.23559  0.22886  0.14103 1.00000 Dx,Dy,Dz 
Cu3 Cu   8 f 0.66231  -0.00254 0.20169 1.00000 Dx,Dy,Dz 
Na1_1 Na   8 f 0.57643  0.02745  0.44987 1.00000 Dx,Dy,Dz 
Na2_1 Na   8 f 0.44742  0.00664  0.14412 1.00000 Dx,Dy,Dz 
O1_1  O    4 e 0.75000  0.14070  0.25000 1.00000 0,Dy,0   
O2_1  O    8 f 0.69621  0.07980  0.40845 1.00000 Dx,Dy,Dz 
O3_1  O    8 f 0.81979  -0.07613 0.45829 1.00000 Dx,Dy,Dz 
O4_1  O    8 f 0.71056  0.89844  0.52811 1.00000 Dx,Dy,Dz 
O5_1  O    8 f 0.34726  0.80441  0.41646 1.00000 Dx,Dy,Dz 
O6_1  O    8 f 0.14434  0.19072  0.31844 1.00000 Dx,Dy,Dz 
O7_1  O    8 f 0.57004  0.86493  0.18793 1.00000 Dx,Dy,Dz 
O8_1  O    8 f 0.00932  0.13618  0.19212 1.00000 Dx,Dy,Dz 
O9_1  O    8 f 0.80949  0.09761  0.57348 1.00000 Dx,Dy,Dz 
O10_1 O    8 f 0.38274  0.84045  0.26494 1.00000 Dx,Dy,Dz 
O11_1 O    8 f 0.12824  0.15154  0.14432 1.00000 Dx,Dy,Dz 
O12_1 O    4 e 0.75000  0.85046  0.25000 1.00000 0,Dy,0   
O13_1 O    8 f -0.07171 0.12542  0.36813 1.00000 Dx,Dy,Dz 
O14_1 O    8 f 0.48655  0.85536  0.43271 1.00000 Dx,Dy,Dz 
S1_1  S    8 f 0.76074  0.00095  0.49084 1.00000 Dx,Dy,Dz 
S2_1  S    8 f 0.41266  0.78545  0.36881 1.00000 Dx,Dy,Dz 
S3_1  S    8 f 0.08644  0.21304  0.21219 1.00000 Dx,Dy,Dz 

loop_
_atom_site_moment.label
_atom_site_moment.crystalaxis_x
_atom_site_moment.crystalaxis_y
_atom_site_moment.crystalaxis_z
_atom_site_moment.symmform
Cu1 -0.241(75) -0.2(1)  -0.757(43)  Mx,My,Mz
Cu2 -0.031(68) -0.256(90)  -0.431(32)  Mx,My,Mz
Cu3 -0.449(35) 0  -0.405(29)  Mx,My,Mz


\end{verbatim}

}
\end{document}